# Influence of Bi substitution with rare-earth elements on the transport properties of BiCuSeO oxyselenides[§]

Andrei Novitskii,[a,b,][*] Illia Serhiienko,[a] Sergey Novikov,[b] Yerzhan Ashim,[c] Mark Zheleznyi,[a] Kirill Kuskov,[a] Daria Pankratova,[a,][‡] Petr Konstantinov,[b] Andrei Voronin,[a] Oleg Tretiakov,[c] Talgat Inerbaev,[d,e] Alexander Burkov,[b] Vladimir Khovaylo[a]

*Abstract:* In this study, we demonstrate that introducing of rare-earth elements, $R$ = La or Pr, into the Bi–O charge reservoir layer of BiCuSeO leads to an increase of both, the charge carrier concentration, and the effective mass. Although the charge carrier mobility slightly decreases upon $Bi^{3+}$ to $R^{3+}$ substitution, the electronic transport properties are significantly improved in a broad temperature range from 100 K to 800 K. In particular, the electrical resistivity decreases by two times, while the Seebeck coefficient drops from 323 µV K$^{-1}$ to 238 µV K$^{-1}$ at 800 K. Thus, a power factor of nearly 3 µW cm$^{-1}$ K$^{-2}$ is achieved for $Bi_{0.92}R_{0.08}CuSeO$ samples at 800 K. Meanwhile, a noticeable decrease of the lattice thermal conductivity is observed for the substituted samples, which can be attributed to the enhanced point defect scattering mostly originated from atomic mass fluctuations between $R$ and Bi. Ultimately, a maximum $zT$ value of nearly 0.34 at 800 K is obtained for the $Bi_{0.92}La_{0.08}CuSeO$ sample, which is ~30% higher than that of pristine BiCuSeO.

## Introduction

The oxyselenides family $MCuSeO$, where $M$ = trivalent cation such as Bi or rare-earth elements, was reported for the first time in 1993 by Boris A. Popovkin group.[1,2] The related compounds (so-called 1111 phases) crystallize in the tetragonal layered ZrCuSiAs structure type with $P4/nmm$ space group and two formula units per cell. The crystal structure is composed of alternately stacked along the $c$ axis insulating $(M_2O_2)^{2+}$ layers with ionic bonds and conducting $(Cu_2Se_2)^{2-}$ layers with covalent bonds, as shown in Fig. 1a.[1,3,4] Generally, the $RCuSeO$ ($R$ = rare-earth) are wide gap $p$-type semiconductors ($E_g$ ~3 eV), with the valence band maximum (VBM) and the conduction band minimum (CBM) at the Γ point of the Brillouin zone.[5–8] In both cases, $RCuSeO$ and BiCuSeO, the VBM is composed of the hybridized Cu 3$d$ and Se 4$p$ orbitals.[9] BiCuSeO is also a $p$-type semiconductor, but with a narrow bandgap of about 0.8 eV. Such evolution in the band structure originates from the presence of Bi 6$p$ orbitals at the bottom of the conduction band.[5,10,11]

To date, BiCuSeO oxyselenides have attracted considerable attention and were intensively studied as promising Pb-free families of thermoelectric materials owing to their intrinsically low thermal conductivity and relatively high Seebeck coefficient. The remarkable thermoelectric performance was reported for dually doped BiCuSeO with the dimensionless figure of merit $zT$ value ~1.5 at 923 K.[12–14] Here $zT$ is defined as $zT = α^2 σ T κ^{-1}$, where $α$, $σ$, $T$ and $κ$ are the Seebeck coefficient, electrical conductivity, absolute temperature, and total thermal conductivity, respectively.[15,16] Considering low $κ$, one of the main approaches of BiCuSeO $zT$ enhancement is the charge carrier concentration optimization, where it can be increased using partial substitution of $Bi^{3+}$ by divalent or monovalent ions inducing extra holes. However, it is challenging to enhance the power factor ($α^2σ$) using the optimization of the carrier concentration since the Seebeck coefficient and the electrical conductivity are inherently coupled through it. Moreover, aliovalent substitution typically leads to deterioration of the charge carrier mobility, which greatly hampers the further possible electrical transport properties improvement.[9,17,18] In recent studies, isovalent substitution has been demonstrated as an effective way to improve the power factor in tetrahedrites, lead chalcogenides, magnesium silicides, and BiCuSeO oxyselenides as well.[19–22] Isovalent substitution is considered as substitution with an element of the same oxidation state as the host atom, which typically does not affect the charge carrier concentration directly, instead it may tune the band and crystal structures along with the introduction of point defects and strain/mass fluctuations. Isovalent substitution in BiCuSeO in particular can lead to several effects, such as an increase in band degeneracy near the Fermi level, tuning of the bandgap, or a decrease in energy offset between the primary and secondary valence bands, which is, in general, beneficial for transport properties.[23–29] In BiCuSeO, isovalent substitution can be realized by substitution with chalcogens on the Se site or by substitution with rare-earth and few other elements (e.g., In or Sb) on the Bi site. According to previous reports, it is expected that rare-earth substitution at the Bi site would lead to the improvement of charge carrier mobility along with the reduction in the charge carrier concentration.[27,28] On the other hand, such a decrease in the carrier concentration can be compensated by using the intensive ball milling during BiCuSeO powder fabrication, which should lead to the introduction of the extra holes in the system *via* Cu vacancies formation:[30,31]

$$Cu_{Cu}^{X} \rightarrow V_{Cu}^{'} + h^{\bullet} + Cu_{surface}^{0}. \quad (1)$$

Among light rare-earth elements (so-called cerium group) La and Pr seem to be the most appropriate elements as isovalent dopants for BiCuSeO since their most stable state in compounds is trivalent in contrast to variable-valence Ce, Sm, or Eu. Hence, herein we have performed a study on the influence of La and Pr substitution at the Bi site in BiCuSeO oxyselenides prepared by a combination of solid-state reaction, intensive ball milling, and densification *via* spark plasma sintering. The experimental study was combined with *ab initio* density functional theory (DFT) calculations in order to clarify the effect of band structure evolution with substitution on transport properties. Overall, our results show that along with some band structure changes the power factor is slightly improved *via* reduced electrical resistivity. Also, isovalent substitution leads to local structural perturbations in the crystal lattice mainly due to mass fluctuation and thus cause thermal conductivity reduction. Consequently, a maximum $zT$ of ~0.34 at 800 K can be achieved for the $Bi_{0.92}La_{0.08}CuSeO$ sample, ~30% enhancement as compared with that of the pristine sample.

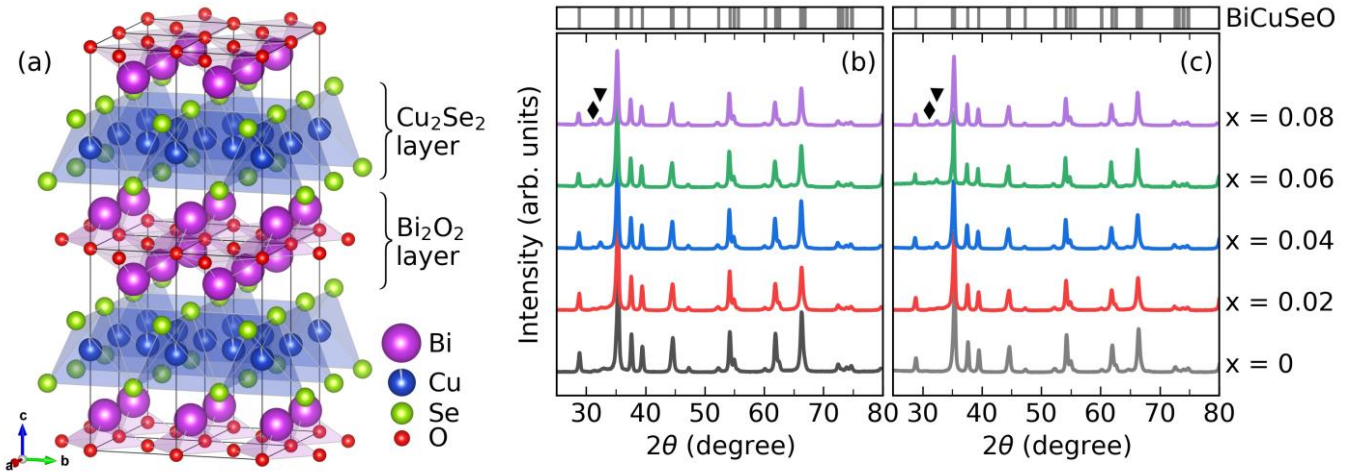

**Figure 1.** (a) Crystal structure of the BiCuSeO and XRD patterns of the (b) $Bi_{1-x}La_xCuSeO$ and (c) $Bi_{1-x}Pr_xCuSeO$ ($x = 0 – 0.08$). Bragg's reflections for the BiCuSeO phase are indicated by grey ticks on the top part of the figure. $Bi_2O_3$ and $Cu_{2-x}Se$ secondary phases are indicated by a black solid triangle (▼) and thin diamond (♦) symbols, respectively. The grey line in Fig. 1c is the XRD pattern for pristine BiCuSeO prepared without ball milling (labeled as BiCuSeO† in the text).

## Experimental details

The starting chemicals for the synthesis of $Bi_{1-x}R_xCuSeO$ ($R$ = La or Pr, $x = 0 – 0.08$) were the fine commercial powders of $Bi_2O_3$ ($\geq$ 99.999%, Reachem), $La_2O_3$ (99.99%, Rare Metallic), $Pr_2O_3$ (99.99%, Shin-Etsu Chemical) and Bi ($\geq$ 99.95%, Component Reaktiv), Se (99.997%, Reachem), Cu ($\geq$ 99.5%, Rushim). The powders were weighed, according to the stoichiometric ratio, and then ball milled (BM) in an argon atmosphere at 400 rpm for 8 hours (planetary micro mill Pulverisette 7 premium line, Fritsch, Germany). The obtained mixture was cold-pressed into pellets and sealed in an evacuated to $10^{-3}$ Torr silica tubes. The tubes were heated to 573 K with the rate of 5 K min$^{-1}$ and held for 8 hours, then cooled to room temperature. The resulting specimens were crushed, ground, and ball-milled again in argon at 400 rpm for 4 hours, cold-pressed, and resealed in another evacuated quartz tube. The samples were annealed at 973 K for 12 hours. The pellets were ball milled one more time in argon at 300 rpm for 8 hours. For ball milling processing, the zirconium oxide vials with a volume of 45 ml and balls with a diameter of 5 mm were used. In order to obtain bulk samples, the powders were densified by spark plasma sintering (Labox 650, Sinter-Land, Japan) at 973 K for 5 minutes under uniaxial pressure of 50 MPa. The cylindrical specimens of 12.7 mm in diameter and 10 mm high were annealed at 973 K for 6 hours in an argon atmosphere. For comparative analysis, one pristine BiCuSeO sample was prepared without using the ball milling during powder preparation, with hand-grinding instead. More details on the experimental procedure can be found in the electronic supporting information file (ESI, Fig. S1).

The synthesized samples were characterized by a range of techniques, including X-ray fluorescence (XRF), powder X-ray diffraction (XRD), scanning electron microscopy (SEM), energy-dispersive X-ray spectroscopy (EDS), Brunauere-Emmette-Teller (BET) surface area analysis, and various physical measurements, such as the electrical and thermal conductivities, Seebeck coefficient and Hall constant measurements. The effective average particle size of the synthesized powders was calculated as $D = 6d^{-1}A^{-1}$ (where $d$ is the density) from the low-temperature adsorption isotherm measurements of the specific surface area, $A$, using a Nova 1200e analyzer (Quantachrome Instruments, USA). The XRF spectra were taken with a ZSK Primus II spectrometer (Rigaku, Japan). X-ray diffraction data were collected over the angular range $20 \leq 2\theta$ (deg) $\leq 80$, on a Miniflex 600-2 diffractometer (Rigaku, Japan), using Co$K_\alpha$ radiation ($\lambda = 1.7903$ Å) at room temperature. Rietveld refinement was performed using PDXL software (Rigaku, Japan). SEM and EDS were performed using a Vega 3SB scanning electron microscope (Tescan, Czech Republic) in conjunction with an EDS detector (x-act, Oxford Instruments, UK). Temperature dependencies of the electrical resistivity and the Seebeck coefficient were simultaneously measured with high resolution (temperature increment ~1 K) by the standard 4-probe and differential methods, respectively, under He atmosphere using a homemade system.[32] The uncertainty of the Seebeck coefficient and the electrical resistivity measurements is 5% ± 0.5 µV/K and 2%, respectively. Thermal conductivity, $\kappa$, was determined from thermal diffusivity measurements using the relationship $\kappa = \chi \cdot C_p \cdot d$, where $\chi$ is the thermal diffusivity coefficient, $C_p$ is the specific heat capacity. The relative bulk density, $d$, was measured by the Archimedes method. Temperature dependencies of the thermal diffusivity were measured by the laser flash diffusivity method (LFA 457 MicroFlash, Netzsch, Germany). In order to minimize errors from the emissivity of the material, the samples were covered by a thin layer of graphite. The thermal diffusivity data were analyzed using a Cowan model with pulse correction.[33] The specific heat $C_p$ was calculated by the Debye model,[34] as is shown in the ESI (Fig. S2). The accuracy in the thermal conductivity values is estimated to be within 8%. All measurements were performed in a direction perpendicular to the SPS pressing direction in the 300 – 800 K temperature range. Temperature-dependent Hall constant was measured by high-resolution double frequency AC technique[35] with a variable magnetic field of 0.15 T (50 Hz) and alternating current (72 Hz) using a laboratory-made installation.[36] The Hall coefficient measurement accuracy is within 5%, along with the total uncertainty of the Hall mobility determination is also 5%. The combined uncertainty for all measurements involved in the $zT$ calculation is less than 15%.

## Computational details

First-principles calculations were carried out using the projector augmented wave (PAW) method[37] and the Perdew-Burke-Ernzerhof (PBE) functional[38] within density functional theory, as implemented in the *ab initio* total energy and molecular dynamics program VASP (Vienna *ab initio* simulation package).[39,40] The generalized-gradient approximation (GGA)[38] of PBE was used for the exchange-correlation functional with the plane wave kinetic energy cut-off of 400 eV. The influence of Bi substitution was examined by constructing a 4×4×2 (256 atoms) supercell using VESTA 3[41] in which five of the Bi atoms were replaced by a La atoms. For both structures, the electron localization function (ELF) was also calculated and used to characterize the degree of electron localization between atoms and the character of the chemical bond.[42] ELF was scaled from 0 to 1 in order to clearly indicate the difference between electron delocalization and full localization. The supercell structures for the mentioned compositions were relaxed using 0.2 Å$^{-1}$ *k*-points spacing in the irreducible Brillouin zone. The Kohn-Sham equations were solved with the denser 0.15 Å$^{-1}$ *k*-points spacing to self-consistently determine the charge distribution of the system's ground state. The non-self-consistent step was performed with 40 *k*-points along high symmetry points to provide accurate electronic band structure. The electronic band structure and density of states plots were produced using the open-source Python package sumo.[43]

**Table 1.** Experimental relative density $d$ and refined lattice constants from Rietveld analysis of X-ray diffraction data for Bi$_{1-x}$R$_x$CuSeO ($R$ = La or Pr, $x$ = 0 – 0.08)

| Specimen | $d$ ± 0.5% (%) | $a$ (Å)   | $c$ (Å)   | $V$ (Å$^3$) |
|----------|----------------|-----------|-----------|-------------|
| $x = 0^\dagger$ | 97.5     | 3.927(1)  | 8.927(8)  | 137.68(3)   |
| $x = 0$  | 94.2           | 3.922(5)  | 8.916(1)  | 137.18(1)   |
| $R$ = La |                |           |           |             |
| $x = 0.02$ | 93.2         | 3.923(4)  | 8.917(3)  | 137.26(7)   |
| $x = 0.04$ | 93.6         | 3.924(3)  | 8.917(8)  | 137.34(9)   |
| $x = 0.06$ | 93.2         | 3.924(3)  | 8.919(1)  | 137.36(1)   |
| $x = 0.08$ | 93.4         | 3.925(2)  | 8.919(7)  | 137.43(9)   |
| $R$ = Pr |                |           |           |             |
| $x = 0.02$ | 93.9         | 3.923(2)  | 8.917(4)  | 137.23(2)   |
| $x = 0.04$ | 93.8         | 3.924(7)  | 8.918(4)  | 137.33(3)   |
| $x = 0.06$ | 94.0         | 3.924(5)  | 8.920(3)  | 137.34(8)   |
| $x = 0.08$ | 94.2         | 3.924(6)  | 8.920(4)  | 137.42(1)   |

## Results and discussion

**Compositional and structural analysis.** Figures 1b, 1c compare the XRD patterns of the SPS-ed and heat-treated Bi$_{1-x}$R$_x$CuSeO ($R$ = La or Pr, $x$ = 0 – 0.08) samples. The XRD pattern of the pristine BiCuSeO specimen prepared from non-milled powder (here and below, labeled as BiCuSeO$^\dagger$) is also presented in Figure 1c (grey line). All the major peaks can be indexed by the BiCuSeO phase (PDF#01-076-6689) with ZrCuSiAs structure-type. The XRD patterns show that a single-phase compound was obtained for the pristine BiCuSeO from non-milled powder. On the other hand, a noticeable amount of impurity phases, Cu$_{2-x}$Se (PDF#01-074-1373) and Bi$_2$O$_3$ (PDF#01-088-2043), in the Bi$_{1-x}$R$_x$CuSeO samples fabricated with ball milling were detected, they account for less than 3 vol.% and 5 vol.%, respectively. The appearance of these impurity phases in the samples with ball milling may be a result of Bi or/and Cu deficiencies formation as was previously reported.[30,44,45] The unit cell parameters for the pristine sample prepared without ball milling are $a$ = 3.92710 Å and $c$ = 8.9278 Å. These values are in good agreement with those reported for samples synthesized by conventional two-step solid-state reaction.[46] However, the ball-milled specimens' cell parameters are noticeably smaller: $a$ = 3.92253 Å and $c$ = 8.9160 Å. Such discrepancy in the lattice constants might result from the fact that the ball-milled BiCuSeO powder before SPS exhibited a smaller effective average particle size of ~400 nm in comparison with ~1 – 10 µm for non-milled as-synthesized BiCuSeO powder, also it could be associated with deficiency formation.[11,47] A peak shift upon the *R*-substitution can be noticed for all samples when $x$ > 0 and can be described by Vegard's law (see

**Table 2.** EDS and XRF results for $Bi_{1-x}R_xCuSeO$ ($R$ = La or Pr, $x = 0 - 0.08$)

| Nominal composition | EDS results | | XRF results | |
|---|---|---|---|---|
| | $(Bi + R)$:Cu:Se ratios | $R/(Bi+R)$ | $(Bi + R)$:Cu:Se ratios | $R/(Bi+R)$ |
| BiCuSeO† | 1.03:0.98:0.99 | 0.00 | 1.02:1.00:1.01 | 0.00 |
| BiCuSeO | 1.04:1.00:0.96 | 0.00 | 0.98:0.99:1.03 | 0.00 |
| $Bi_{0.98}La_{0.02}CuSeO$ | 1.04:1.01:0.95 | 0.01 | 0.99:1.00:0.98 | 0.02 |
| $Bi_{0.96}La_{0.04}CuSeO$ | 1.02:1.03:0.95 | 0.04 | 0.99:0.96:1.05 | 0.03 |
| $Bi_{0.94}La_{0.06}CuSeO$ | 1.03:1.02:0.95 | 0.06 | 1.00:1.00:1.04 | 0.05 |
| $Bi_{0.92}La_{0.08}CuSeO$ | 1.02:1.01:0.97 | 0.07 | 0.97:0.97:1.03 | 0.07 |
| $Bi_{0.98}Pr_{0.02}CuSeO$ | 1.04:1.00:0.96 | 0.02 | 0.99:0.97:1.01 | 0.02 |
| $Bi_{0.96}Pr_{0.04}CuSeO$ | 1.03:1.00:0.97 | 0.03 | 0.99:0.96:1.02 | 0.03 |
| $Bi_{0.94}Pr_{0.06}CuSeO$ | 1.03:1.02:0.95 | 0.05 | 0.98:0.97:1.03 | 0.05 |
| $Bi_{0.92}Pr_{0.08}CuSeO$ | 1.04:0.99:0.97 | 0.06 | 0.99:0.98:1.03 | 0.06 |

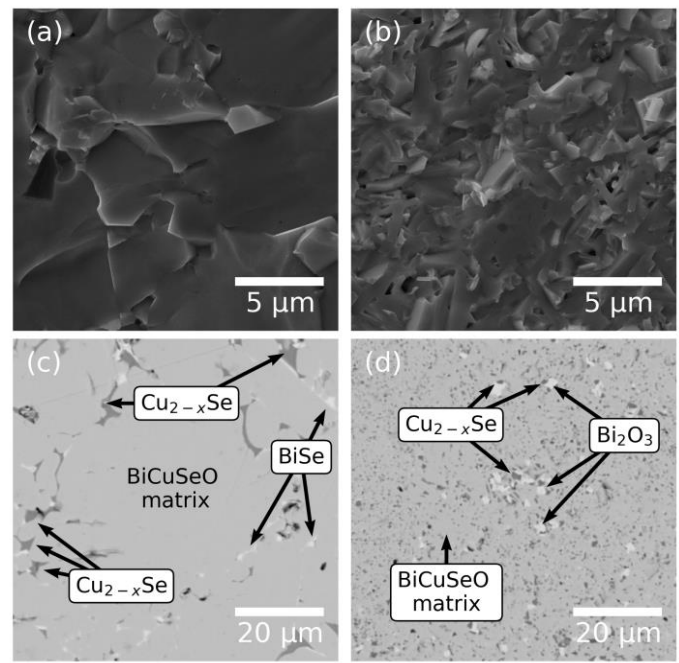

**Figure 2.** SEM micrographs of the (a, b) fracture surfaces and (c, d) polished surfaces for BiCuSeO† and BiCuSeO bulk samples, respectively. The dark areas are $Cu_{2-x}Se$ phases, and the light regions are BiSe (c) and $Bi_2O_3$ (d) phases.

Table 1).[48] The increase of the cell parameters with $R$ content can be explained by a larger ionic radius of $R^{3+}$ (1.03 Å for $La^{3+}$ and 1.01 Å for $Pr^{3+}$, respectively) as compared to $Bi^{3+}$ (0.96 Å).[49] The Rietveld refinement results are summarized in Table 1 (all the Rietveld refinements can be found in the ESI, Fig. S3). EDS analysis was made for all samples to compare the effective and nominal La and Pr concentrations. It was revealed that for some samples the actual $R$ content is slightly lower than the nominal values (Table 2). Moreover, EDS analysis also detected traces of BiSe and $Cu_{2-x}Se$ even for BiCuSeO sample prepared without ball milling (see Figs. 2c, 2d, and Figs. S4 – S7). Nevertheless, the relative elemental proportions of $(Bi + R)$:Cu:Se found from the EDS and XRF spectra are in good agreement with each other (see Table 2).

SEM images of the fractured cross-section of the densified BiCuSeO† and BiCuSeO samples are displayed in Figure 2. All the samples exhibited a similar lath-like microstructure with randomly arranged platelet grains stacked densely (Fig. S8), typical for the BiCuSeO based compounds.[4,9,10] The relative density was measured to be above 93% of theoretical density (see Table 1). However, there is a noticeable difference in the average grain size, which is 1 – 20 microns in thickness for BiCuSeO† and 300 – 500 nm in thickness for BiCuSeO, respectively. The fine microstructure presented in Figure 2b is also representative of all $Bi_{1-x}R_xCuSeO$ ($R$ = La or Pr, $x = 0 - 0.08$) samples. Smaller grain size for the ball-milled samples is attributed to the use of a combination of intensive ball milling and rapid densification *via* SPS.

**Table 3.** Room-temperature electrical transport properties of $Bi_{1-x}R_x CuSeO$ ($R$ = La or Pr, $x$ = 0 – 0.08) samples; $m^*$ were calculated within single parabolic band model with acoustic phonon scattering (SPB-APS)

| Nominal composition | $\sigma$ ($\Omega^{-1}$ cm$^{-1}$) | $\alpha$ ($\mu$V K$^{-1}$) | $p_H$ ($10^{18}$ cm$^{-3}$) | $\mu_H$ (cm$^2$ V$^{-1}$ s$^{-1}$) | $m^*$ ($m_e$) |
|---|---|---|---|---|---|
| BiCuSeO without BM[30] | 0.2 | 58 | 2.7·10$^{-1}$ | 10.10 | – |
| BiCuSeO 500 min BM[30] | ~2 | 373 | 5.2·10$^0$ | 2.77 | – |
| BiCuSeO[†] | 0.7 | 62 | 1.1·10$^0$ | 8.81 | – |
| BiCuSeO | 46.6 | 266 | 8.2·10$^1$ | 3.41 | 1.76 |
| $Bi_{0.98}La_{0.02}CuSeO$ | 51.0 | 243 | 1.1·10$^2$ | 3.06 | 2.59 |
| $Bi_{0.96}La_{0.04}CuSeO$ | 61.5 | 213 | 1.6·10$^2$ | 2.48 | 2.72 |
| $Bi_{0.94}La_{0.06}CuSeO$ | 81.2 | 186 | 2.0·10$^2$ | 2.33 | 2.77 |
| $Bi_{0.92}La_{0.08}CuSeO$ | 76.8 | 173 | 2.3·10$^2$ | 2.22 | 2.95 |
| $Bi_{0.98}Pr_{0.02}CuSeO$ | 53.8 | 232 | 1.1·10$^2$ | 3.05 | 2.66 |
| $Bi_{0.96}Pr_{0.04}CuSeO$ | 61.2 | 212 | 1.7·10$^2$ | 2.27 | 2.75 |
| $Bi_{0.94}Pr_{0.06}CuSeO$ | 69.1 | 173 | 1.9·10$^2$ | 2.25 | 2.85 |
| $Bi_{0.92}Pr_{0.08}CuSeO$ | 93.4 | 169 | 2.5·10$^2$ | 2.24 | 3.01 |

**Electrical transport.** The room-temperature electrical transport data are summarized in Table 3. Hall coefficient measurements indicate $p$-type conduction with Hall charge carrier concentrations of $1.1 \times 10^{18}$ and $8.2 \times 10^{19}$ cm$^{-3}$ for the pristine samples of BiCuSeO[†] and BiCuSeO, respectively. The carrier concentration increases to ~$2.5 \times 10^{20}$ cm$^{-3}$ for substituted samples with $x$ = 0.08 (see Table 3). The Hall charge carrier concentration $p_H$ is calculated as $p_H = 1/R_H e$, where $R_H$ is the Hall coefficient, and $e$ is the elementary charge.

A dramatic increase of charge carrier concentration of ball-milled BiCuSeO for more than one order of magnitude (see Table 3) originates from the introduction of acceptor crystal defects (mainly Cu vacancies) during ball milling, as was previously reported by L.-D. Zhao *et al.*[9] With the substitution of La or Pr at the Bi site, $p_H$ further increases in the whole temperature range (Fig. 3a). Moreover, the carrier concentration of all the samples increased with increasing temperature due to the thermal excitation. The charge carrier density in BiCuSeO may be affected by various crystal defects, such as anti-site defects,[50] deviations from nominal oxidation states[44,51] or vacancies[47,52] as it was previously reported for oxychalcogenide-based compounds.[53–56] Although the role of defects in this system requires further detailed studies, it is clear that the enhanced electrical conductivity of the ball-milled BiCuSeO and that of the substituted samples should be attributed to the increased charge carrier density (see Table 3).

The Hall mobility, $\mu_H$, is calculated from the Hall coefficient, $R_H$, and the electrical conductivity, $\sigma$ as $\mu_H = \sigma R_H$. Generally, in materials with multiple scattering mechanisms affecting the carrier mean free path, the temperature dependence of charge carrier mobility is complex and total $\mu$ can be evaluated by Matthiessen's rule, which assumes scattering channels are independent of each other: $\mu^{-1} = \Sigma_i \mu_i^{-1}$, where $\mu_i$ represents the mobility of a specified scattering mechanism. However, according to the temperature dependence of the Hall mobility fitted by a power law $\mu \propto T^r$, the value of $r$ is close to –1.9 (see Fig. 3c) indicating that the acoustic phonon scattering is the dominant scattering mechanism at moderate temperatures. The present results are in general agreement with the literature data.[25,26,57,58] In contrast, at low temperatures, the failure of a single power-law to fit the mobility data is likely due to the presence of multiple scattering mechanisms. The decrease of $\mu_H$ upon $R$-substitution at low temperatures can be attributed to increased ionized impurity scattering, which is typically more significant at lower temperatures.[59]

For a single parabolic band semiconductor where acoustic phonons dominate the scattering of carriers, the Hall mobility can be represented by[59,60]

$$\mu_H \propto \frac{C_{ll}}{T^{3/2} m_I^* \left(m_b^*\right)^{3/2} \Delta_{def}^2}, \quad (2)$$

where $C_{ll}$ is the elastic constant for longitudinal vibrations, $m_I^*$ is the inertial effective mass, $m_b^*$ is the density-of-states effective mass of a single band ($m_d^* = N_V^{2/3} m_b^*$, with $m_d^*$ is the density-of-states effective mass, $N_V$ is the band degeneracy; for single parabolic band $m_d^* = m_b^* = m_I^*$), and $\Delta_{def}$ is the deformation potential characterizing the carrier-phonon interaction. Observed $r < -1.5$ (see Fig. 3c) can be related to the temperature dependence of $C_{ll}$, $m^*$ and $\Delta_{def}$ or the presence of polar optical scattering.[25]

Figures 3b and 3d show the room-temperature variation of $p_H$ and $\mu_H$ data from this study with substitution fraction on the Bi site, theoretical nominal $p_H$ for substitution with 1$^+$ and 2$^+$ valence elements ($M^{1+}$ and $M^{2+}$ substitution), the minimum carrier mobility, $\mu_{min}$, for $M^{1+}$ and $M^{2+}$ substitution and data reported in previous studies.[46,61–66] $\mu_{min}$ was estimated by the Ioffe-Regel criterion $\mu_{min} = 0.33 ea/(\hbar p^{1/3})$, where $p$ and $a$ are nominal carrier concentration and lattice parameter.[67] The $R$-substituted samples prepared with ball milling possess similar charge carrier concentration values as compared with Na-, Sr-, Ca- and Ba-substituted samples (Fig. 3b). However, it is much lower than the charge carrier density obtained for Pb-substituted samples and considerably higher than that in Mg-substituted ones. The mobility of all the studied samples is lower than 5 cm$^2$ V$^{-1}$ s$^{-1}$, i.e., close to the Ioffe-Regel limit, indicating a strong scattering of charge carriers (Fig. 3d).

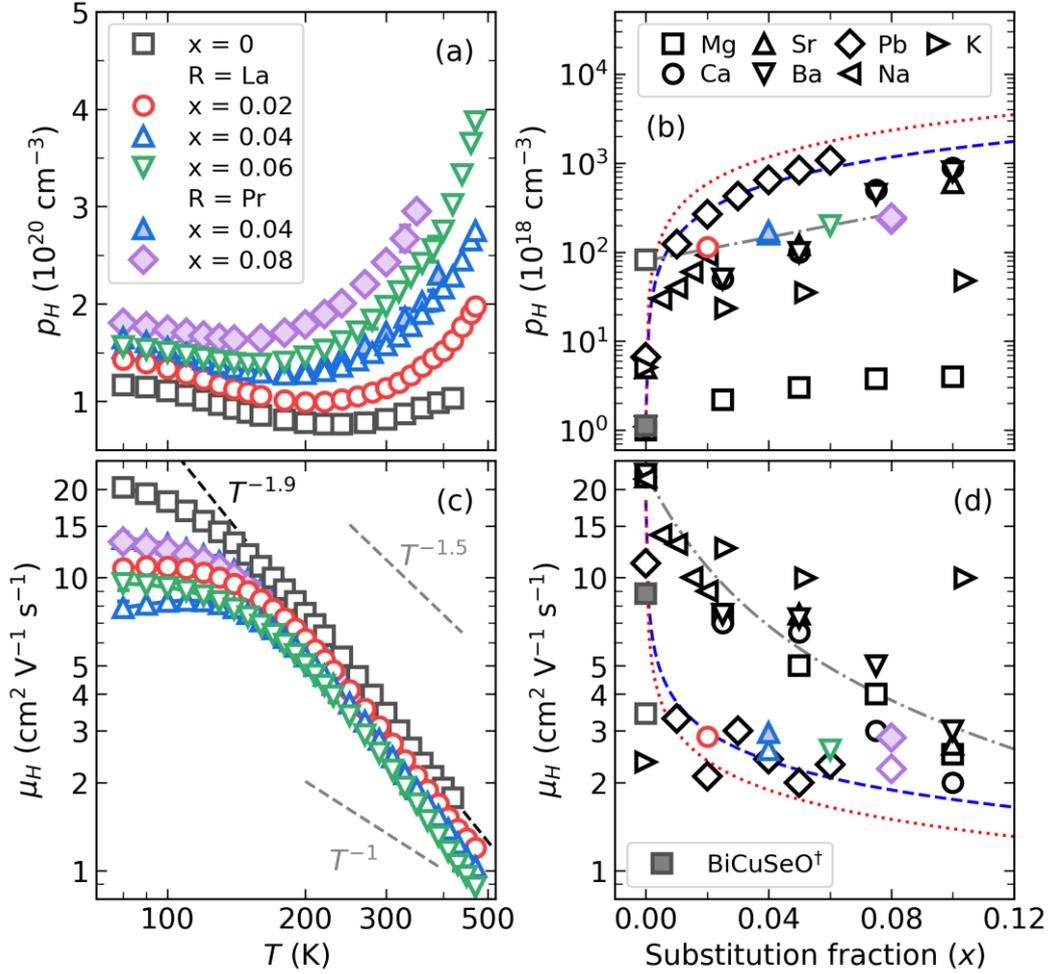

**Figure 3.** Temperature dependence of the (a) Hall carrier concentration and (c) Hall carrier mobility (log-log scale) for the series $Bi_{1-x}R_xCuSeO$ ($R$ = La or Pr, $x = 0 - 0.08$). The dashed lines are the $\mu \propto T^r$ relations. (b) Hall carrier concentration as a function of substitution fraction at Bi site with additional data from the literature.[11,46,61,63–66] The dashed and dotted lines are calculated nominal carrier concentration: the blue is for $M^{1+}$ substitution, the red is for $M^{2+}$ substitution, the grey one is a guide to the eye. (d) Hall carrier mobility as a function of substitution fraction at Bi site with additional data from the literature.[46,61,63–66] The dashed and dotted lines are estimated by the Ioffe-Regel criterion the minimum carrier mobility, $\mu_{min}$: the blue is for $M^{1+}$ substitution, the red is for $M^{2+}$ substitution; the grey dash dotted line is a guide to the eye.

In fact, such changes in the charge carrier concentration and mobility are not expected from the simple electron counting since $Bi^{3+}$ is substituted by the isovalent $R^{3+}$ ($R$ = La or Pr). On the one hand, compared with the pristine BiCuSeO, the electronic states at Fermi energy in $Bi_{0.92}La_{0.08}CuSeO$ are inclined towards higher energy, implying that more electronic states are available and thus the concentration of charge carriers is induced by substitution (see Fig. 4). Moreover, as was speculatively mentioned before the charge carrier density in BiCuSeO can be to some extent also affected by crystal defects and especially Cu vacancies, which formation can be induced upon substitution at the Bi site as it was previously reported for Pb-substituted oxytellurides.[53] At the same time, the charge carries in BiCuSeO are predominantly scattered by acoustic phonons and thus their mobility is closely related to the effective mass and the carrier-phonon coupling as shown in Eq. (2).[59,60] The effective mass, $m^*$, is largely affected by the ionicity of the corresponding chemical bond and considering the noticeable difference in the electronegativity between Bi and rare-earth elements ($\chi_{Bi}$ ~2.02, $\chi_{La}$ ~1.10 and $\chi_{Pr}$ ~1.13)[68] can be increased due to the increased ionicity of the $R$–O bond.[69,70]

In turn, for degenerate semiconductors, assuming parabolic band structure and acoustic phonon scattering approximation, the effective mass can be roughly estimated by the well-known relation:[60]

$$\alpha = \frac{8\pi^2 k_B^2 T}{3eh^2} m_S^* \left(\frac{\pi}{3p_H}\right)^{2/3}, \quad (3)$$

where $\alpha$ is the Seebeck coefficient, $k_B$ is the Boltzmann constant, $h$ is the Planck's constant, $m_S^* \approx m_d^*$. The effective mass was estimated from the slope of the $\alpha \cdot p_H^{2/3}$ vs. $T$ plot curves (see Fig. 5) and presented in Table 3. Although the utilization of this relation may introduce some errors arising from non-parabolicity and multi-valley conduction (one must keep in mind that this DOS effective mass is a mixed effective mass of multiple bands since there are at least two valence bands close to the top of the VBM), it can still be used to describe the trend of $m_d^*$ with $R$-substitution. As shown in Figure 5, the $\alpha \cdot p_H^{2/3}$ vs. $T$ dependencies are linear for all samples at $T < 400$ K. The estimated DOS effective masses of the $R$-substituted BiCuSeO compounds are higher than those of the pristine compound. As was suggested above, such a trend can originate from the electronegativity difference of the $R$ and Bi atoms. Ionicity of the Bi–O bond ($\Delta\chi$ ~1.4, Pauling ionicity ~0.40) is significantly lower than that of the $R$–O bond ($\Delta\chi$ ~2.3, Pauling ionicity ~0.75). Thus, $R$-substitution significantly increases the bond ionicity of the $[Bi_2O_2]^{2+}$ layers and leads to an increased carrier effective mass as well as reduced charge carrier mobility according to Eq. (2).

This results in good consistency with the DFT calculations, which shows that for the substituted samples the contribution of the heavy band increases (Fig. 4) despite the fact that La-derived states contribute substantially only to the energy regions far from the top of the valence band with La $5d$ and La $4f$ states located at –15 eV and 2.7 eV, respectively (Figs. S9, S10). Moreover, according to the non-parabolic fitting of the band edges (for 3 $k$-points) the estimated hole effective mass at VBM increases with substitution from $0.6m_e$ for BiCuSeO to $1.1m_e$ for 8% La-substituted BiCuSeO. The band structure calculations carried out for pristine BiCuSeO confirm the indirect bandgap with VBM on the Γ–M line and CBM located at the Γ point of the Brillouin zone, which also is in a good agreement with the previous reports.[5]

An analysis of the electron localization function distribution (Fig. 6) revealed that additional La atoms at Bi site in the unit cell decrease the ELF between La and O atoms, which can be interpreted as the increase of La–O bond ionicity (Fig. 6b) in comparison with Bi–O (Fig. 6a). The isosurface values between Bi and O atoms are approximately 0.4 (Fig. 6a) and decrease to ~0.25 when the Bi atom is replaced by La (Fig. 6b). In all other aspects, the ELF distribution is similar to those reported for pristine BiCuSeO and LaCuSeO: the charges of these systems are mainly located around Bi, La, Se, and O atoms, while Cu atoms are weakly bonded with other atoms.[71,72] Thus, in general, replacing Bi with La increases the overall ionic character of the system with the noticeable broadening of the bandgap from 0.28 eV for pristine BiCuSeO to 0.40 eV for La-substituted sample, respectively (Figs. 4, 6). It should be noted that the calculated band gaps are somehow underestimated, which is typical for DFT-GGA calculations.[73] However, the change of $E_g$ with $R$-substitution still can be evaluated since we focus on the difference between band gaps of similar compounds. Besides, the bandgap changes from indirect to direct with the VBM and CBM at the Γ point of the Brillouin zone, which can be expected for $R$CuSeO.[5,10]

The temperature dependence of the transport properties for the BiCuSeO† sample with some reference data is presented in the ESI (Fig. S11). Since the charge carrier concentration of semiconductors is sensitive to crystal defects, it dramatically increases due to introduced acceptor defects during the ball milling process (see Table 3).[30] The specific electrical conductivity of acceptor semiconductors can be expressed as $\sigma = ep\mu$, where $p$ is the hole concentration, and $\mu$ is the hole mobility; thus, the enhancement of the electrical conductivity is mainly connected with the increase of the carrier concentration (Fig. S11a). Moreover, the temperature behavior of $\sigma$ is changed to a regime of the degenerate semiconductor after ball milling. Considering the layered features of the BiCuSeO structure, the anisotropic transport behavior can be expected (Fig. S12).[30,74] However, the ratios of the electrical resistivity and the Seebeck coefficient measured in parallel and perpendicular directions to the applied pressure direction during the SPS process do not exceed 5%, so the electrical transport properties for BiCuSeO polycrystalline specimens understudy can be assumed to be isotropic (Fig. S13). The electronic transport properties were measured during heating-cooling cycles and the measurement results were completely reproducible. The specific electrical resistivity and the Seebeck coefficient decrease almost linearly with increasing La or Pr content due to the higher charge carrier concentration (see Table 3). Moreover, $\rho$ and the absolute values of $\alpha$ increase with temperature for all the samples, as expected for heavily doped semiconductors (Figs. 7a, 7b). The effect of thermally activated minority carriers can be observed at high temperatures as resistivity and the magnitude of the Seebeck coefficient begin to decrease. For the pristine BiCuSeO, the extrinsic-intrinsic transition temperature ranges from 670 to 700 K, while it is shifted to a higher temperature for $R$-substituted specimens, suggesting that the increased value of the bandgap is high enough to prevent the bipolar conduction (Fig. 4b). Due to the strong trade-off between the electrical resistivity and the Seebeck coefficient, the power factor, $PF = \alpha^2 \rho^{-1}$, is only slightly affected by the substitution. The maximum $PF$ value of $(3.1 \pm 0.3)$ μW cm$^{-1}$ K$^{-2}$ was achieved for Bi$_{0.92}$R$_{0.08}$CuSeO at 800 K, which is ~35% higher than that for the pristine BiCuSeO (see Fig. 7c).

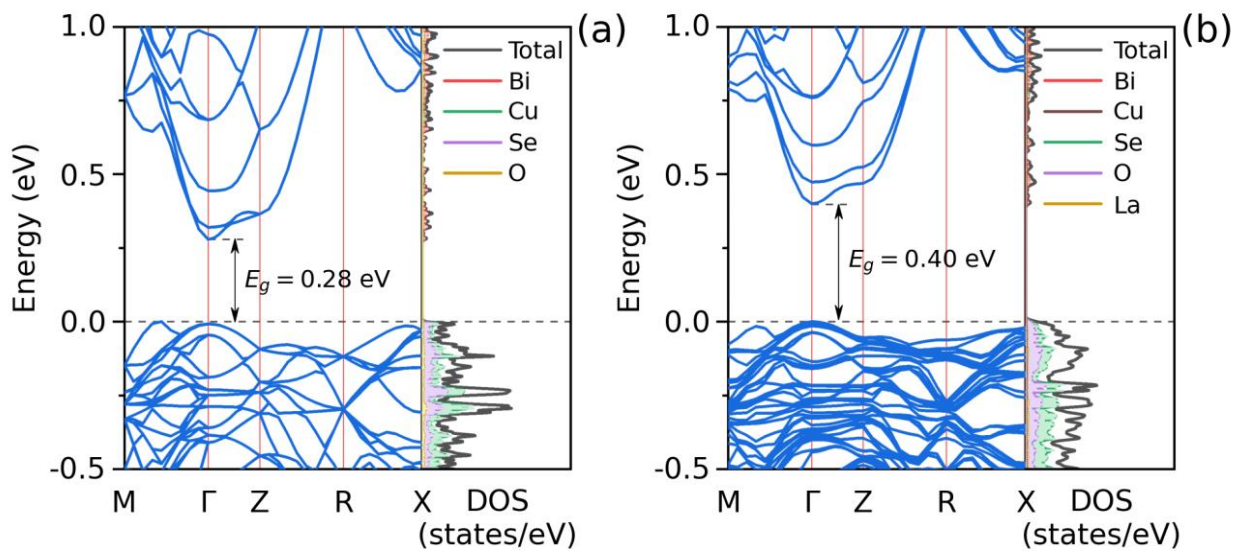

**Figure 4.** Electronic band structure and projected DOS of (a) BiCuSeO and (b) $Bi_{0.92}La_{0.08}CuSeO$. Fermi level is referred to the top of the valence band (dashed lines at 0 eV).

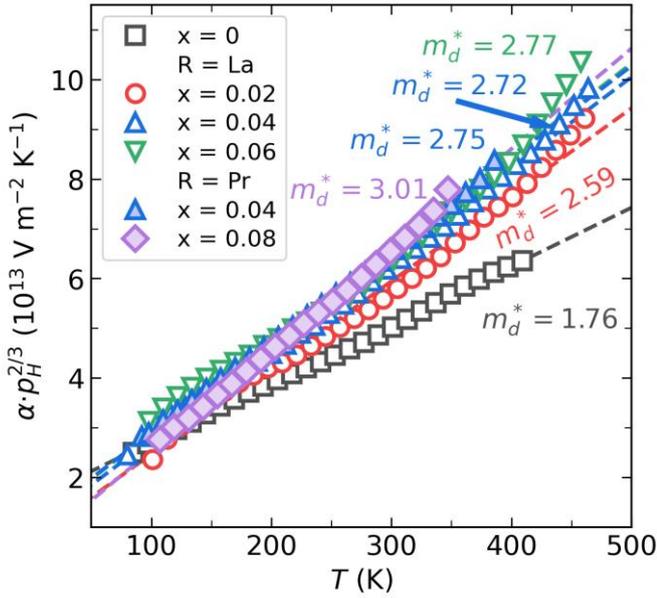

**Figure 5.** The plot of $\alpha \cdot p^{2/3}$ as a function of $T$ for the estimation of the DOS effective masses based on Eq. (3) for Bi$_{1-x}$R$_x$CuSeO ($R$ = La or Pr, $x = 0 - 0.08$) samples.

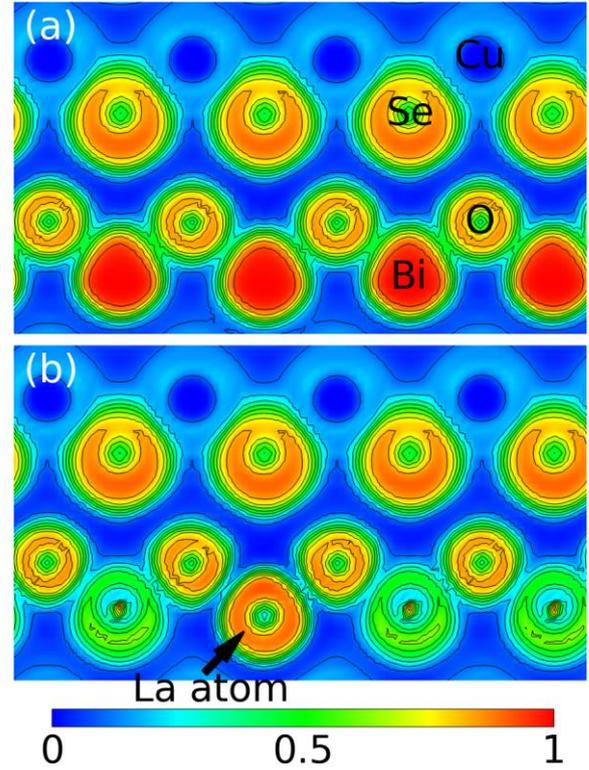

**Figure 6.** The map distribution of the calculated electron localization function in (a) Bi$_{64}$Cu$_{64}$Se$_{64}$O$_{64}$ and (b) Bi$_{59}$La$_5$Cu$_{64}$Se$_{64}$O$_{64}$ with a cut by the (100) plane. Blue color (ELF = 0) represents no electron localization, green (ELF = 0.5) represents homogeneous electron distribution and red (ELF = 1) corresponds to full electron localization.

**Thermal transport.** The thermal conductivity measurements were performed along both parallel and perpendicular to the pressing directions. It is necessary to mention that the difference between parallel and perpendicular measurements is found to be about 15 – 20% (Fig. S13). We found that the values of the thermal conductivity measured in the parallel direction are lower compared with those determined in the perpendicular direction. It is commonly assumed, that polycrystalline samples, produced from powders by conventional consolidation techniques are isotropic. Usually, measurements of the thermal conductivity with the flash technique are carried out in the direction parallel to the pressing direction, while measurements of the electrical transport properties are performed in the perpendicular direction. However, our results indicate that for the oxyselenides-based compounds, which are intrinsically anisotropic, layered materials, taking anisotropy into account is essential even for the pressed polycrystalline samples. It can be suggested that the observed anisotropy in the thermal conductivity is mainly related to a preferential grain growth along the perpendicular to the applied pressure direction during the SPS process (see Figs. S11, S12, and S13), which is in a good accord with the literature data[52] and was confirmed by SEM studies (Fig. S12). Hereinafter, only measurements carried out in the perpendicular direction are presented.

A small decrease in the total thermal conductivity of ~10% is observed for the BiCuSeO after ball milling at room temperature (Fig. 8a). The reduction in the thermal conductivity after ball milling is most likely attributed to the grain boundary scattering and the point defect scattering due to the refined grains (Fig. 2). Further substitution with Pr and La reduces the room-temperature thermal conductivity by ~15%. However, the reduction becomes less pronounced at higher temperatures where all the samples exhibit $\kappa$ ~0.7 W m$^{-1}$ K$^{-1}$. Since the electrical resistivity decreases with the substitution of $R$ at the Bi site, the electronic contribution to the thermal conductivity becomes more significant. The total thermal conductivity is a sum of the lattice, $\kappa_{lat}$, and the electronic, $\kappa_{el}$, thermal conductivities, where the electronic contribution can be calculated through Wiedemann–Franz law, $\kappa_{el} = L\sigma T$, where $L$ is the Lorenz number.[60] The values of $\kappa_{lat}$ are calculated from $\kappa_{lat} = \kappa_{tot} - \kappa_{el}$. Lorenz number calculations are based on the assumption of acoustic phonon scattering and a single parabolic band, the details can be found in the ESI (Fig. S14a). As expected, the ball milling process and further substitution result in the reduction of the lattice thermal conductivity at room temperature approximately by 30% for Bi$_{0.92}$R$_{0.08}$CuSeO. For all the samples, the temperature dependence of $\kappa_{lat}$ obeys the $\kappa_{lat}$ ~$T^{-1}$ law, indicating a substantial phonon-phonon scattering in the whole temperature range. The electronic contribution, $\kappa_{el}$, is less than 9% of the total thermal conductivity for the heavily substituted samples and less than 3% for the pristine ones. Overall, the lattice thermal conductivity for all the samples reaches ~0.62 W m$^{-1}$ K$^{-1}$ at high temperatures, which is close to the theoretical minimum thermal conductivity ($\kappa_{min}$ for BiCuSeO ~0.59 W m$^{-1}$ K$^{-1}$) calculated by Cahill's model (Fig. 8b).[75] Observed behavior of $\kappa_{lat}$ as well as other transport properties are in good agreement with the previously reported results (Fig. S15).[30,62,76]

For a better understanding of the thermal conductivity mechanism in BiCuSeO, the experimental data are fitted and analyzed by the Debye–Callaway model,[77] which can be expressed as:

$$\kappa_{lat} = \frac{k_B}{2\pi^2 v_a} \cdot \left(\frac{k_B T}{\hbar}\right)^3 \cdot \int_0^{\theta_D/T} \tau_c \frac{x^4 e^x}{(e^x - 1)^2} dx, \quad (4)$$

where $v_a$ is the average velocity of sound, $\theta_D$ is the Debye temperature, $x = \hbar\omega/k_B T$, $\hbar$ is the reduced Planck's constant, $\omega$ is the phonon frequency, and $\tau_c$ is the total phonon relaxation time. The overall relaxation rate $\tau_c^{-1}$ can be determined by combining the various scattering processes based on Matthiessen's rule:

$$\tau_c^{-1} = \sum_i \tau_i^{-1} = \tau_U^{-1} + \tau_{PD}^{-1} + \tau_{GB}^{-1}, \quad (5)$$

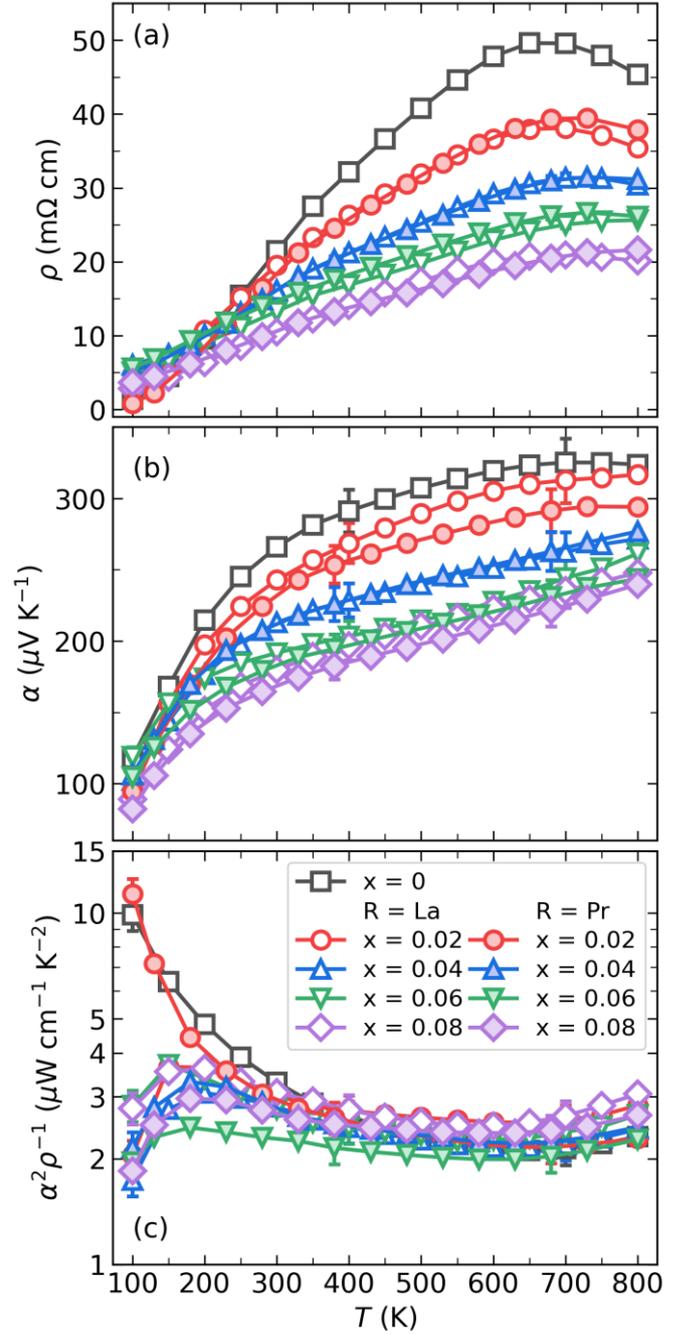

**Figure 7.** Temperature dependence of (a) the electrical resistivity, (b) the Seebeck coefficient and (c) the power factor for Bi$_{1-x}$R$_x$CuSeO ($R$ = La or Pr, $x$ = 0 – 0.08) samples.

where $\tau_U$, $\tau_{PD}$, and $\tau_{GB}$ are the relaxation times for Umklapp scattering, point defect scattering, and grain boundary scattering respectively. Umklapp relaxation rate is defined as:

$$\tau_U^{-1} = B\omega^\alpha \left(\frac{T}{\theta_D}\right)^\beta \exp\left(-\frac{\theta_D}{bT}\right), \quad (6)$$

where $\alpha$, $\beta$, and $b$ are constants,[78] that were chosen to be 2, 1, and 3 respectively to provide the best fit to the experimental data. Point defect term is expressed as:

$$\tau_{PD}^{-1} = A\omega^4 = \frac{V_a}{4\pi v_a^3}(\Gamma + \Phi)\omega^4, \quad (7)$$

where $V_a$ is the average atomic volume, $\Gamma$ is the disorder scattering parameter, and $\Phi$ is the vacancies scattering parameter.[79] Finally,

$$\tau_{GB}^{-1} = \frac{v_a}{L}, \quad (8)$$

where $L$ is the average grain size.

The $\tau_U$ could be extracted from the data for the pristine sample, assuming that the sound velocity, Debye temperature, and Umklapp process relaxation time are independent of substitution. The disorder scattering parameter $\Gamma$ includes both, the strain field $\Gamma_S$ and the mass fluctuation scattering $\Gamma_M$, and can be calculated by the model of Abeles and Slack as $\Gamma = \Gamma_M + \Gamma_S$ (for details see ESI).[80,81] $\Phi$ scattering parameter can be calculated using similar to $\Gamma_M$ expression including broken-bond term, as was proposed by R. Gurunathan *et al.*[82]

$$\Gamma_M = \frac{\sum_{i=1}^{n} c_i \left(\frac{\overline{M}_i}{\overline{\overline{M}}}\right)^2 f_i^1 f_i^2 \left(\frac{M_i^1 - M_i^2}{\overline{M}_i}\right)^2}{\sum_{i=1}^{n} c_i}, \quad (9)$$

$$\Gamma_S = \frac{\sum_{i=1}^{n} c_i \left(\frac{\overline{M}_i}{\overline{\overline{M}}}\right)^2 f_i^1 f_i^2 \varepsilon_i \left(\frac{r_i^1 - r_i^2}{\overline{r}_i}\right)^2}{\sum_{i=1}^{n} c_i}, \quad (10)$$

where $n$ is the number of different crystallographic sublattice types in the lattice and $c_i$ is the relative degeneracy of the respective sites. In the pure BiCuSeO, $n = 4$, $c_i = 2$, $\overline{\overline{M}}$ is the average atomic mass, $f_i^k$ is the fractional occupation of the $k$-th atom on the $i$-th site, $M_i^k$ and $r_i^k$ are the atomic mass and radius of the $k$-th atom, $\varepsilon_i$ is a function of the Grüneisen parameter, $\gamma$, which characterizes the anharmonicity of the lattice,[80] and for BiCuSeO is calculated to be 114. $\overline{M}_i$ and $\overline{r}_i$ are the average atomic mass and radius on the $i$-th site, respectively:

$$\overline{M}_i = \sum_k f_i^k M_i^k, \quad \overline{r}_i = \sum_k f_i^k r_i^k. \quad (11)$$

Combining all the mentioned scattering mechanisms the calculated lattice thermal conductivities are shown in Fig. 8b (solid lines). As expected from the large mass difference between Bi, La, and Pr (208.98, 138.91, and 140.91, respectively) along with the size difference, the introduction of $R$ dopants increases the mass and strain fluctuations. Thus, point defect scattering plays an important role in decreasing the thermal conductivity. Further, the Umklapp scattering mechanism seems to dominate the scattering of heat-carrying phonons at high-temperature. However, there is a distinct difference of ~10% between the experimental data and the theoretical simulation at $T > 600$ K, especially for the samples with $x > 0.04$. Thus, additional scattering mechanisms such as normal processes scattering, or electron-phonon scattering should be also considered for model optimization.

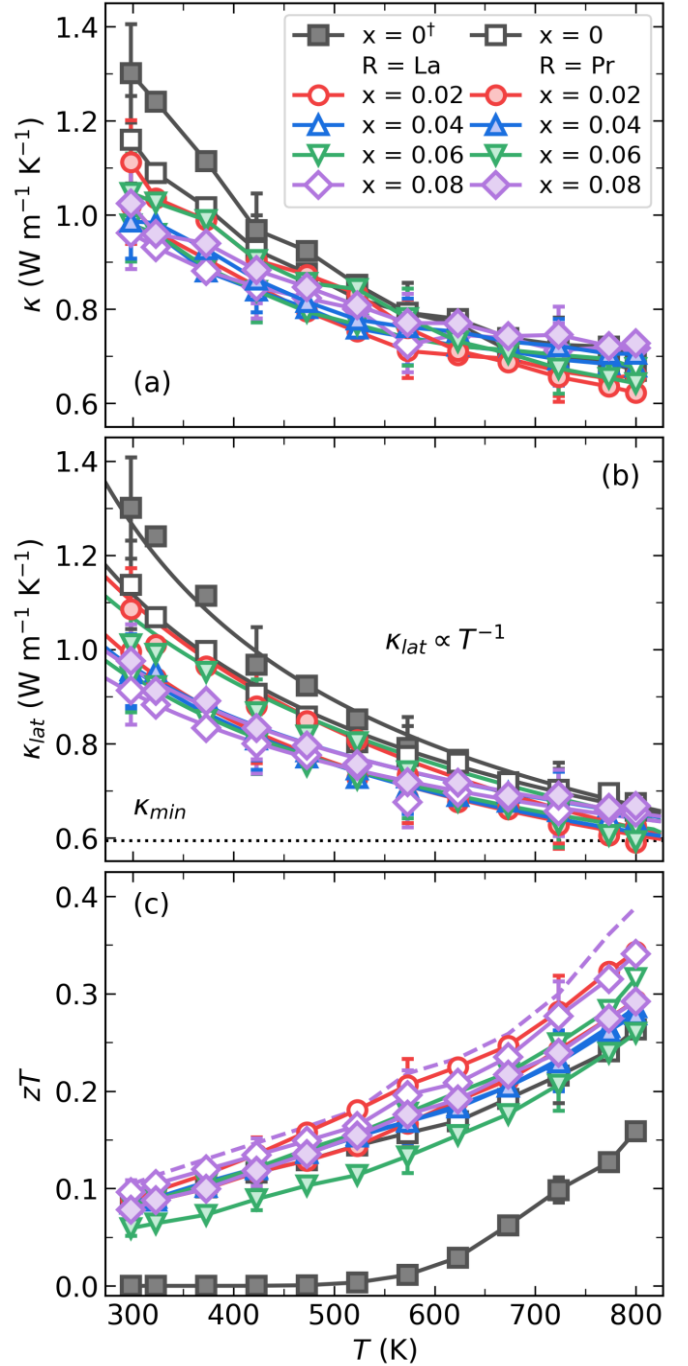

**Figure 8.** Temperature dependence of (a) the total thermal conductivity, (b) the lattice contribution to the thermal conductivity and (c) the figure of merit for Bi$_{1-x}$R$_x$CuSeO ($R$ = La or Pr, $x = 0 – 0.08$) and BiCuSeO[†] samples. Solid lines in (b) are calculated by the Debye–Callaway model. The minimum thermal conductivity is calculated by the Cahill's model.[75]

**Thermoelectric efficiency.** The thermoelectric figure of merit $zT$ is shown as a function of temperature in Fig. 8c. The values of $zT$ are calculated from the measured electrical and thermal transport properties. $R$-substitution on the Bi site enhances the power factor, mainly due to the increase of the carrier concentration. The maximum value of $zT$ is found to be 0.34 ± 0.04 at 800 K for the Bi$_{0.92}$La$_{0.08}$CuSeO sample, ~31% and ~70% enhancements as compared with those of the pristine BiCuSeO samples, prepared with and without the ball milling, respectively. Also shown in Fig. 8c is the $zT$ dotted curve calculated for Bi$_{0.92}$La$_{0.08}$CuSeO using the thermal conductivity values measured in a parallel direction to the pressing direction. Consideration of the thermal conductivity anisotropy leads to a ~10% reduction in $zT$ for BiCuSeO based compounds. The $zT$ values shown here are consistent with those presented for pristine polycrystalline samples and are lower than those of other alkali or alkaline-earth metals substituted BiCuSeO.[30,46,63–65] While the charge carrier concentration obtained for the *R*-substituted samples in this study and for Ba-, Na-, K- and Sr-substituted BiCuSeO is similar,[63] the decrease in the charge carrier mobility along with slightly affected thermal conductivity upon the *R*-substitution may explain the observed difference in the $zT$ value.

## Conclusions

Polycrystalline Bi$_{1-x}$R$_x$CuSeO (*R* = La or Pr, $0 \le x \le 0.08$) samples were synthesized by combining a solid-state reaction, ball milling and SPS. The phase composition and microstructure were examined, and the related thermoelectric transport properties were investigated. The results indicated that the dependence of the charge carrier concentration on composition in the *p*-type Bi$_{1-x}$R$_x$CuSeO samples does not agree with the expected carrier density obtained from the simple electron counting and may be attributed to the band structure tuning, which was also observed in previous reports.[27,28] Another mechanism that is mainly speculative, but should be mentioned as possible is the formation of crystal defects, such as anti-sites, vacancies, etc., also resulting in an enhancement of $p_H$. Therefore, the electrical resistivity and the Seebeck coefficient both moderately decreased, and the power factor of ~3.1 μW cm$^{-1}$ K$^{-2}$ was achieved for Bi$_{0.92}$La$_{0.08}$CuSeO at 800 K. In addition, *R*-substitution at the Bi site increases mass and strain fluctuations, and thus leads to a noticeable decrease of the lattice thermal conductivity. However, at high temperatures, where the lattice thermal conductivity is limited by the phonon-phonon scattering and the electronic thermal conductivity became more significant, the total thermal conductivity exhibited almost the same value for all the samples close to 0.7 W m$^{-1}$ K$^{-1}$. As a result, a maximum $zT$ value of ~0.34 was achieved for the Bi$_{0.92}$La$_{0.08}$CuSeO specimen at 800 K.

## CRediT authorship contribution statement

**Andrei Novitskii:** Conceptualization, Methodology, Formal analysis, Investigation, Writing – original draft, Writing – Review & Editing, Visualization. **Illia Serhiienko:** Conceptualization, Methodology, Investigation. **Sergey Novikov:** Investigation. **Yerzhan Ashim:** Investigation. **Mark Zheleznyi:** Investigation. **Kirill Kuskov:** Investigation. **Daria Pankratova:** Investigation. **Petr Konstantinov:** Investigation. **Andrei Voronin:** Resources, Project administration, Funding acquisition. **Oleg Tretiakov:** Resources. **Talgat Inerbaev:** Resources, Writing – Review & Editing. **Alexander Burkov:** Resources, Data curation, Writing – Review & Editing. **Vladimir Khovaylo:** Resources, Writing – Review & Editing, Supervision.

## Conflicts of interest

The authors declare no competing financial interest.

## Acknowledgements

The study was carried out with financial support from the Russian Science Foundation (project No. 19-79-10282). The calculations were performed at the Cherry supercomputer cluster provided by the Materials Modeling and Development Laboratory at NUST "MISIS" (supported *via* the Grant from the Ministry of Education and Science of the Russian Federation No. 14.Y26.31.0005). A. Novitskii is very grateful to Professor S. Kobeleva from National University of Science and Technology MISIS and to P. Korotaev from Dukhov Research Institute of Automatics for their helpful and stimulative discussions.

## Data availability

The data that support the findings of this study are openly available in Mendeley Data repository at http://doi.org/10.17632/nhgk9hhtmp.1.

# Influence of Bi substitution with rare-earth elements on the transport properties of BiCuSeO oxyselenides

Andrei Novitskii,[a,b,*] Illia Serhiienko,[a] Sergey Novikov,[b] Yerzhan Ashim,[c] Mark Zheleznyi,[a] Kirill Kuskov,[a] Daria Pankratova,[a,‡] Petr Konstantinov,[b] Andrei Voronin,[a] Oleg Tretiakov,[c] Talgat Inerbaev,[d,e] Alexander Burkov,[b] Vladimir Khovaylo[a]

[a] National University of Science and Technology MISIS, 119049 Moscow, Russia

[b] Ioffe Institute, 194021 St. Petersburg, Russia

[c] The University of New South Wales, 2052 Sydney, Australia

[d] Sobolev Institute of Geology and Mineralogy, 630090 Novosibirsk, Russia

[e] L.N. Gumilyov Eurasian National University, 010000 Nur-Sultan, Kazakhstan

[‡] Present address: Luleå University of Technology, 97187 Luleå, Sweden

[*] E-mail: novitskiy@misis.ru

**Experimental details**

Ball-milling was carried out using a Pulverisette 7 planetary micro mill (Fritsch, Germany) with zirconium oxide balls (diameter of 5 mm, powder-to-ball ratio of 1:5) and vials (45 ml) in an argon atmosphere in reverse mode. To avoid overheating of the powders, the milling periods never exceeded 5 minutes. For instance, ball milling for 8 hours was performed as 96 cycles of "5 min milling + 5 min break". In order to obtain bulk samples, the ball-milled powders were put into cylindrical graphite die with a diameter of 12.7 mm and compressed in an SPS machine at room temperature for 1 minute under a uniaxial pressure of 50 MPa in an evacuated to ~10 Pa chamber, which then was filled with Ar. The temperature of the samples was gradually raised to a sintering temperature of 973 K with a heating rate of 50 K min$^{-1}$; after the dwelling for 5 min at a sintering temperature, the pressure was gradually reduced to ~10 MPa, and the sample was cooled to room temperature with a cooling rate of 20 K min$^{-1}$. Figure S1 displays a schematic illustration of the synthesis route used in this work. For comparative analysis, one pristine BiCuSeO sample was prepared without using the ball milling (without BM-1, BM-2, and BM-3 steps, respectively; see Fig. S1) with hand-grinding instead. All the densified disk-shaped specimens had a dimension of 12.7 mm diameter × 10 mm height, respectively. Obtained bulks were annealed at 973 K for 6 hours in an argon atmosphere.

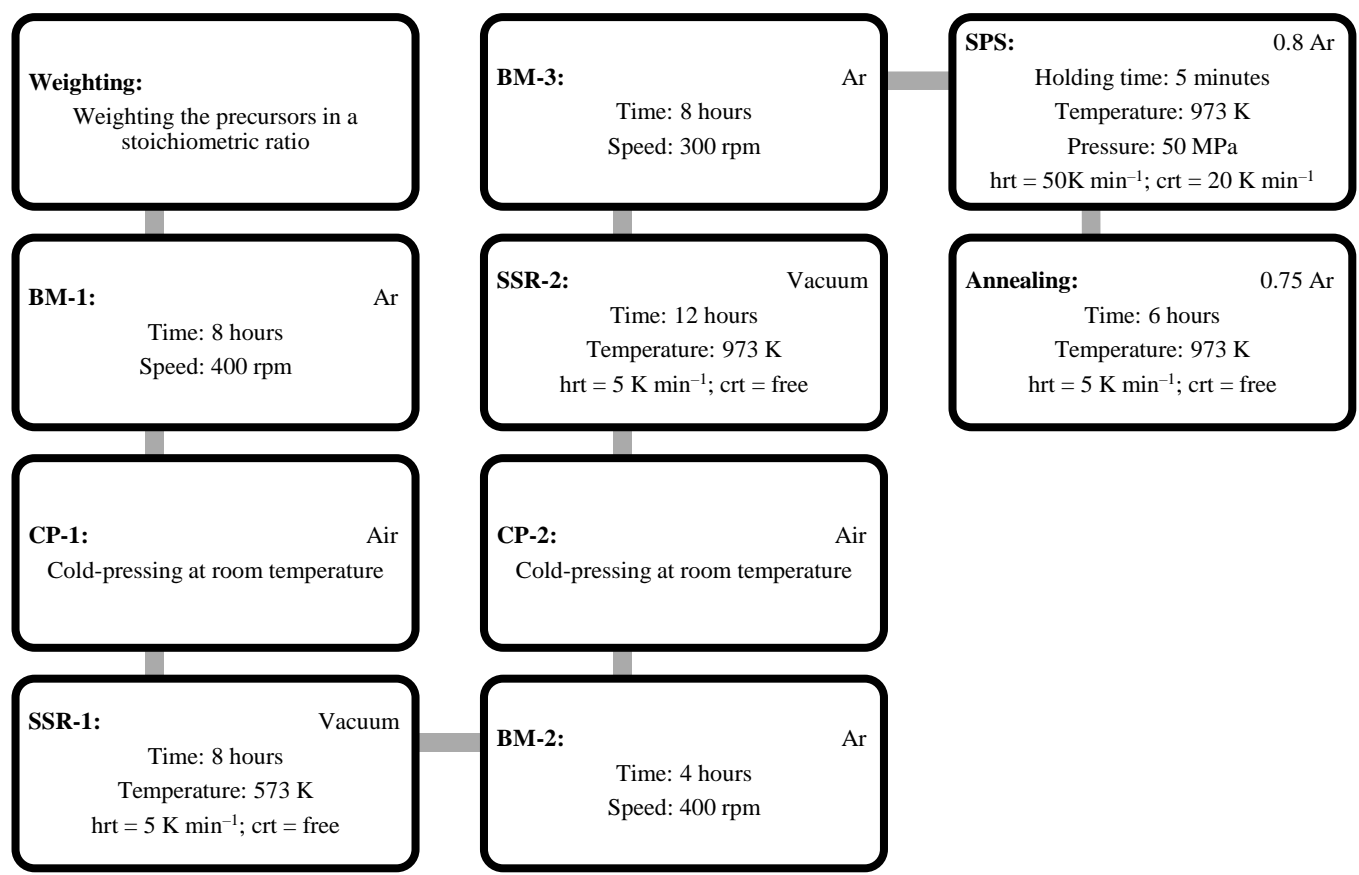

**Figure S1.** Schematic illustration of the fabrication route for the Bi$_{1-x}$R$_x$CuSeO (R = La or Pr, $x$ = 0, 0.02, 0.04, 0.06, 0.08) samples (hrt and crt are heating and cooling rates, respectively).

## Thermal diffusivity and heat capacity

The thermal diffusivity ($\chi$) was measured by a laser flash method (LFA 457 MicroFlash, Netzsch, Germany) under a continuous Ar flow, and the data is shown in Fig. S2a. Generally, experimental data for the specific heat capacity of BiCuSeO oxyselenides can be fitted and analyzed by the Debye model as it was shown by many reports (see Fig. S2b).[1,2] In this work, the heat capacity was calculated by the Debye model (similar to the Dulong-Petit law when $T > \theta_D$):[3,4]

$$C_p = \gamma T + \frac{9Rn}{M} \cdot \left(\frac{T}{\theta_D}\right)^3 \cdot \int_0^{\theta_D/T} \frac{x^4 e^x}{(e^x - 1)^2} dx, \tag{S1}$$

where $\gamma$ is the Sommerfeld coefficient, $R$ is the ideal gas constant, $n$ is the number of atoms, $M$ is the molar mass of a primitive cell, $\theta_D$ is the Debye temperature, and $x = h\nu/k_B T$. The Debye temperature and the Sommerfeld coefficient were assumed to be independent of chemical composition; $\theta_D = 243$ K and $\gamma = 3.075 \cdot 10^{-7}$ J g$^{-1}$ K$^{-2}$ as reported by C. Barreteau and L. Pan.[2,5] The first term of Equation (S1) corresponds to the electronic part, while the second one corresponds to the phonon contribution to the $C_p$. For undoped BiCuSeO the $C_p$ approaches the Dulong-Petit limit at high temperature ($C_p = 3nR/M$; 0.271 J·g$^{-1}$·K$^{-1}$ for BiCuSeO). However, it can be expected that due to a significant difference in atomic mass between La, Pr, and Bi (138.91, 140.91, and 208.98, respectively) the heat capacity of the $R$-doped samples should be slightly higher than for the undoped sample. All calculations were carried out considering this mass difference (see Fig. S2c).

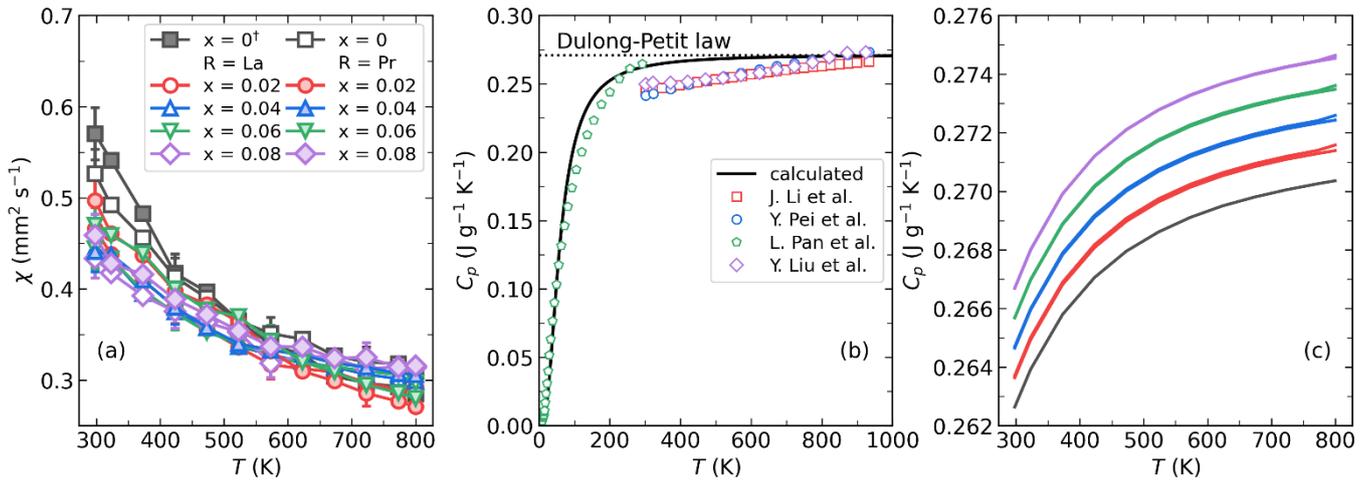

**Figure S2.** Temperature dependence of (a) the thermal diffusivity, $\chi$, for Bi$_{1-x}$R$_x$CuSeO ($R$ = La or Pr, $x$ = 0 – 0.08), (b) the specific hear capacity calculated by Debye model (black solid line) and experimental data from previous reports[5–8] (adapted from Ref.[9]) and (c) the calculated heat capacity (solid lines) for doped Bi$_{1-x}$R$_x$CuSeO ($R$ = La or Pr, $x$ = 0 – 0.08) used for thermal conductivity calculation.

## Rietveld refinement

The final refinement was carried out assuming a tetragonal symmetry with a space group of *P4/nmm* and taking the pseudo-Voigt function for the peak profiles. Figure S3 displays the observed (circles) and fitted (solid red lines) diffraction patterns taken at room temperature, with their differences plotted below the XRD patterns (solid blue lines). They agree very well with $R_p \leq 8.6\%$, $R_{wp} \leq 10.9\%$, and $\chi^2 \leq 2.3$.

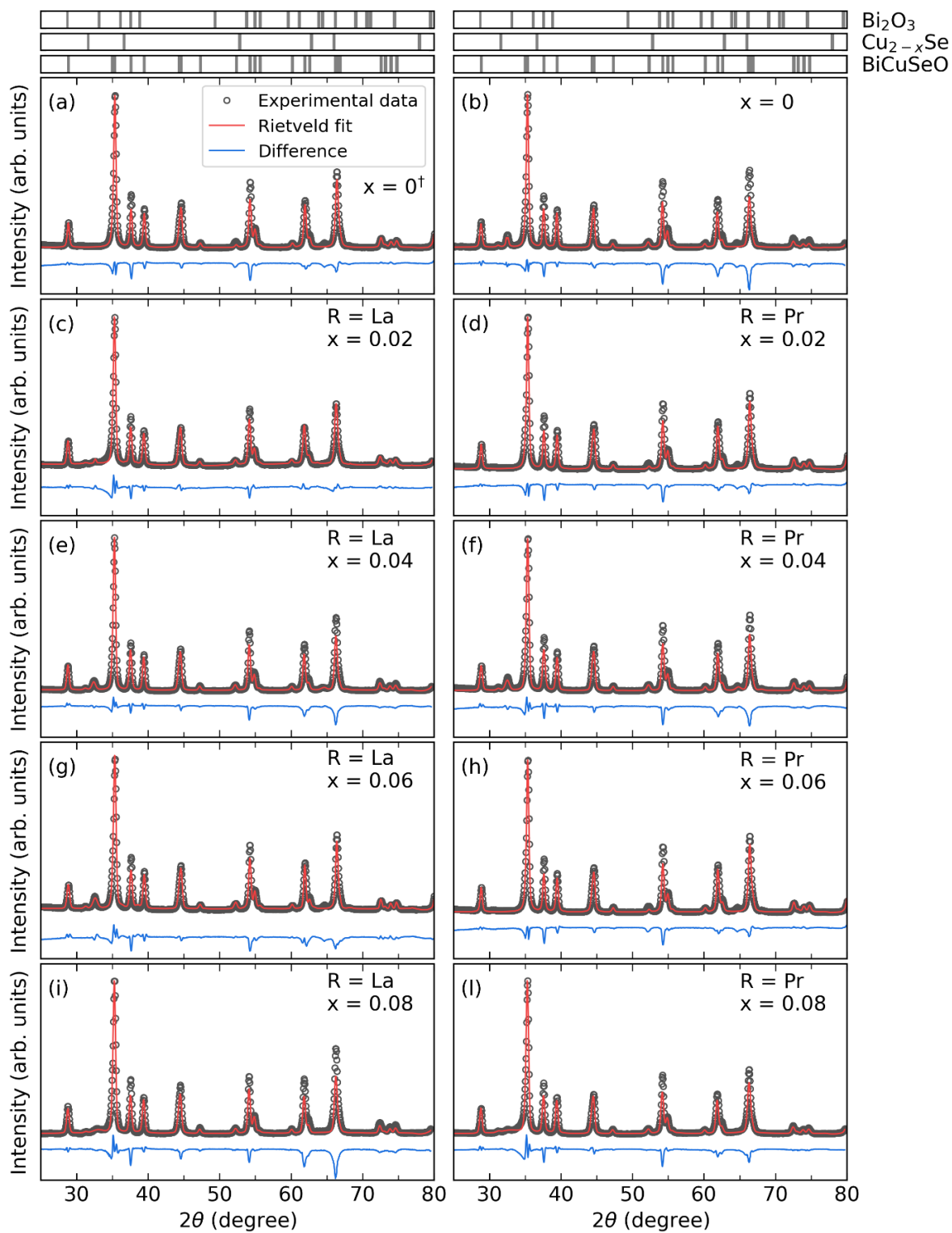

**Figure S3.** Rietveld refinement of powder XRD pattern for BiCuSeO[†] and $Bi_{1-x}R_x$CuSeO ($R$ = La or Pr, $x = 0 - 0.08$) specimens.

## Compositional and structural analyses

### Energy-dispersive X-ray spectroscopy.

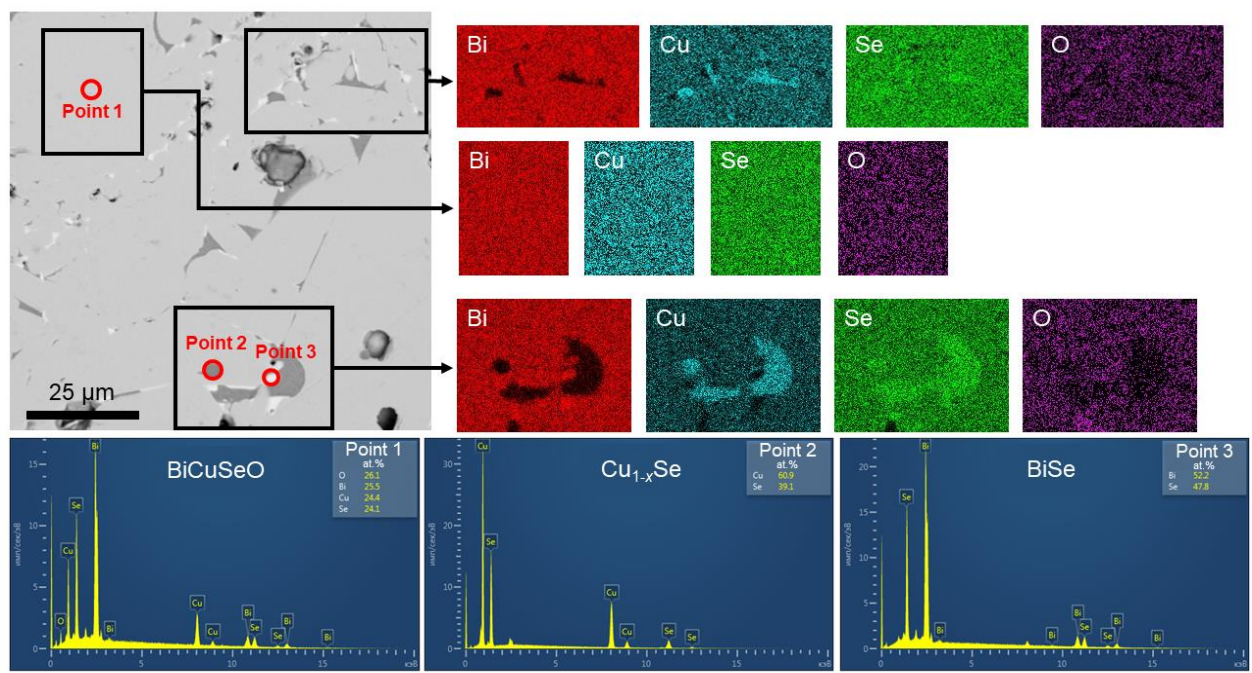

**Figure S4.** SEM image of the polished surface, EDS spectra, and mapping for pristine BiCuSeO bulk prepared from non-milled powder (hereinafter labeled as BiCuSeO[†]).

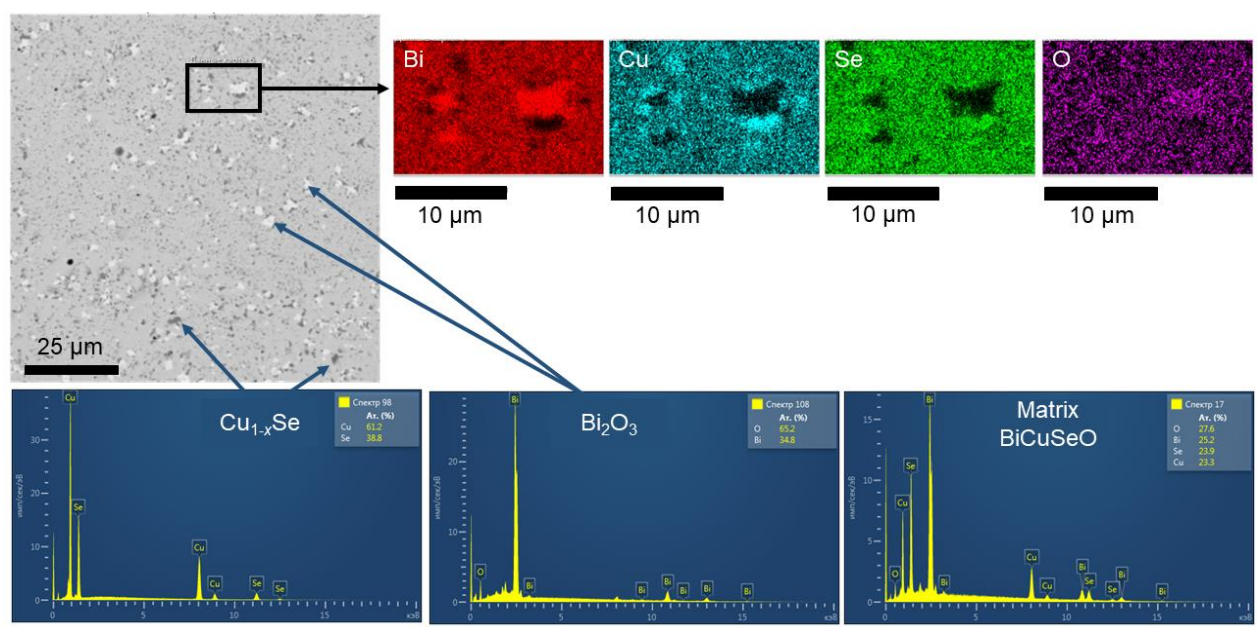

**Figure S5.** SEM image of the polished surface, EDS spectra, and mapping for pristine BiCuSeO bulk prepared from ball milled powder.

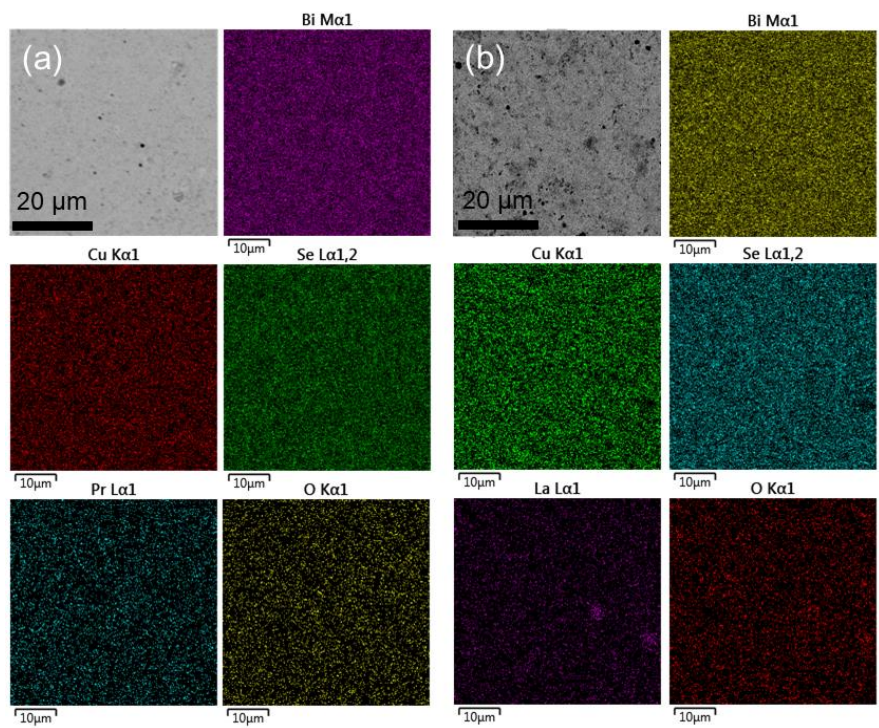

**Figure S6.** SEM micrographs of the polished surfaces and corresponding EDS mapping for (a) $Bi_{0.96}Pr_{0.04}CuSeO$ and (b) $Bi_{0.96}La_{0.04}CuSeO$.

**X-ray diffraction and X-ray fluorescence.** XRD pattern evolution for the starting mixture with a nominal composition of BiCuSeO during the solid-state fabrication route followed by SPS is shown in Figure S7a. The elemental composition of the studied $Bi_{1-x}R_xCuSeO$ ($R$ = La or Pr, $0 \leq x \leq 0.08$) samples were examined by X-ray fluorescence (XRF), the XRF spectra were taken with a ZSK Primus 9 II spectrometer (Rigaku, Japan), employing the standard setup at room temperature. The XRF spectra for some samples are shown in Figure S7b. The relative elemental proportion of (Bi + $R$):Cu:Se:O determined from XRF and EDS can be found in the main text (see Table 2).

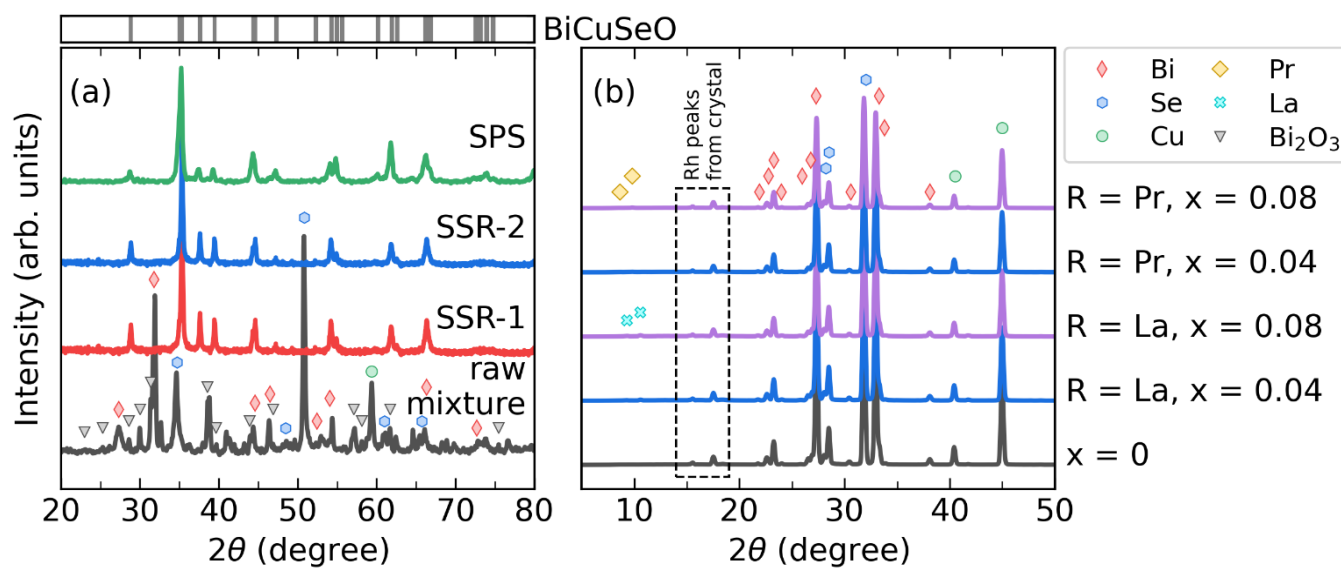

**Figure S7.** (a) XRD patterns for raw mixture, powders after first (SSR-1) and second (SSR-2) steps of solid-state reaction, after spark plasma sintering for BiCuSeO; (b) XRF spectra for the $Bi_{1-x}R_xCuSeO$ ($R$ = La or Pr, $x$ = 0, 0.04, 0.08) samples.

**Scanning electron microscopy.** SEM images of the fractured cross-section of the densified BiCuSeO†, BiCuSeO, Bi$_{0.96}$Pr$_{0.04}$CuSeO, and Bi$_{0.96}$La$_{0.04}$CuSeO samples are displayed in Figure S5. All the samples exhibited a similar lath-like microstructure with randomly arranged platelet grains stacked densely, which is typical for BiCuSeO based compounds. The grain size after SPS remained 5 – 20 μm for BiCuSeO† and 200 – 700 nm for Bi$_{1-x}$R$_x$CuSeO ($R$ = La or Pr, $x$ = 0, 0.04, 0.08).

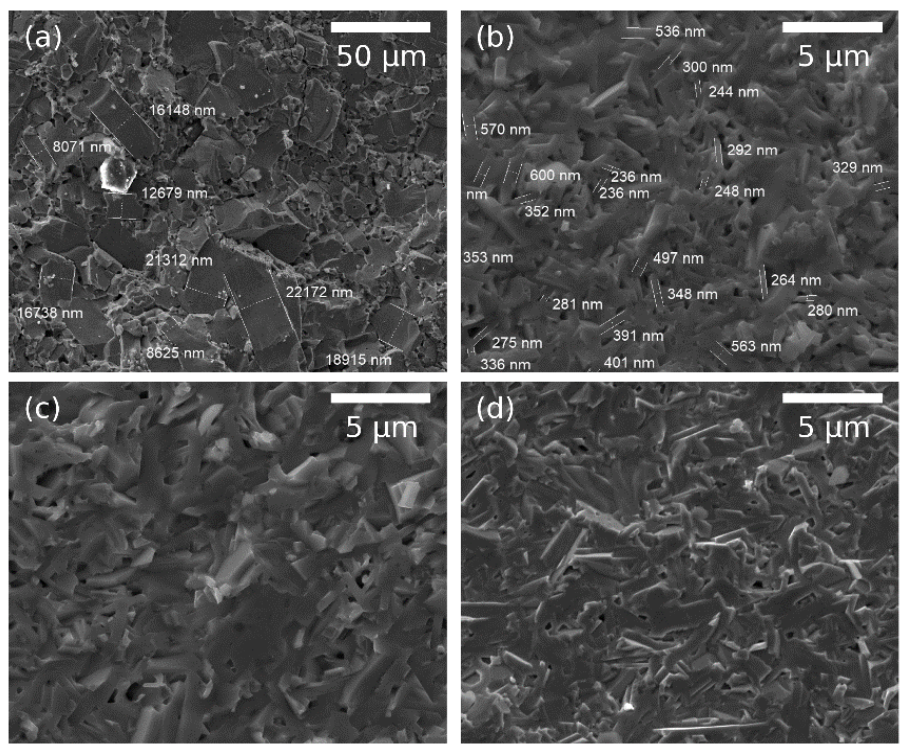

**Figure S8.** SEM micrographs of the fracture surfaces for (a) BiCuSeO†, (b) BiCuSeO, (c) Bi$_{0.96}$Pr$_{0.04}$CuSeO, and (d) Bi$_{0.96}$La$_{0.04}$CuSeO bulk sample.

## First principles calculations

**The total and partial DOS.** The computed projected density-of-states for $Bi_{59}La_5Cu_{64}Se_{64}O_{64}$ are in good agreement with the results for LaCuSeO reported by H. Hiramatsu *et al.*[10]

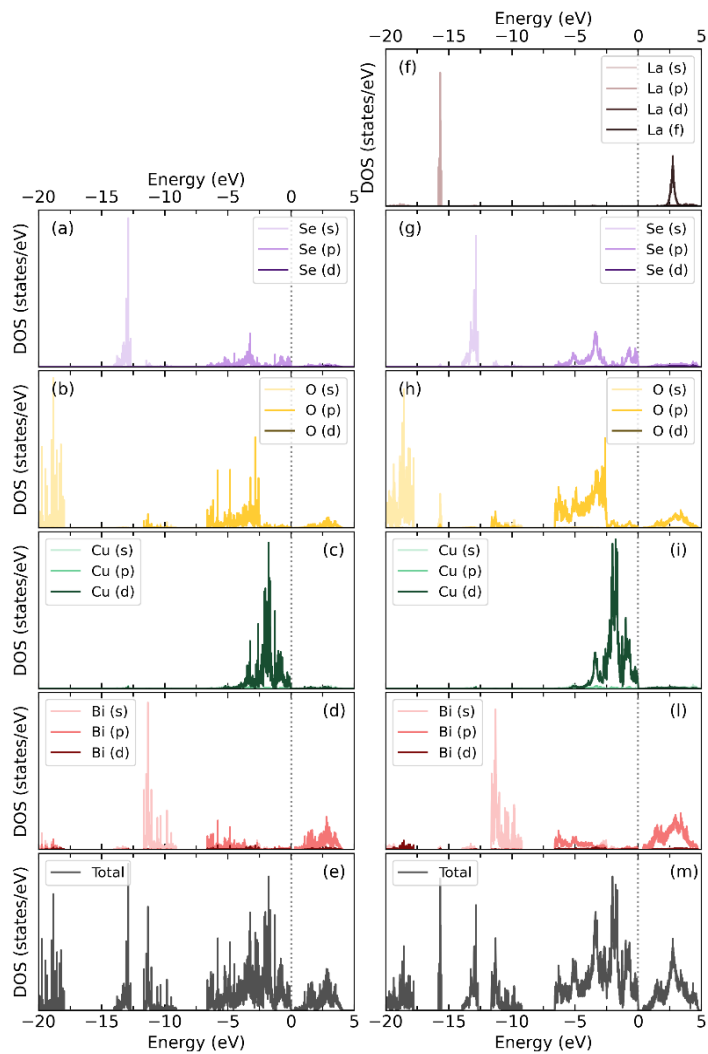

**Figure S9.** The total DOS and partial DOSs of (a) $Bi_{64}Cu_{64}Se_{64}O_{64}$ and (b) $Bi_{59}La_5Cu_{64}Se_{64}O_{64}$.

**Projected band structures.** Projected band structures, where the contributions of Bi *p*, Cu *d*, Se *p* to a band are plotted.[11]

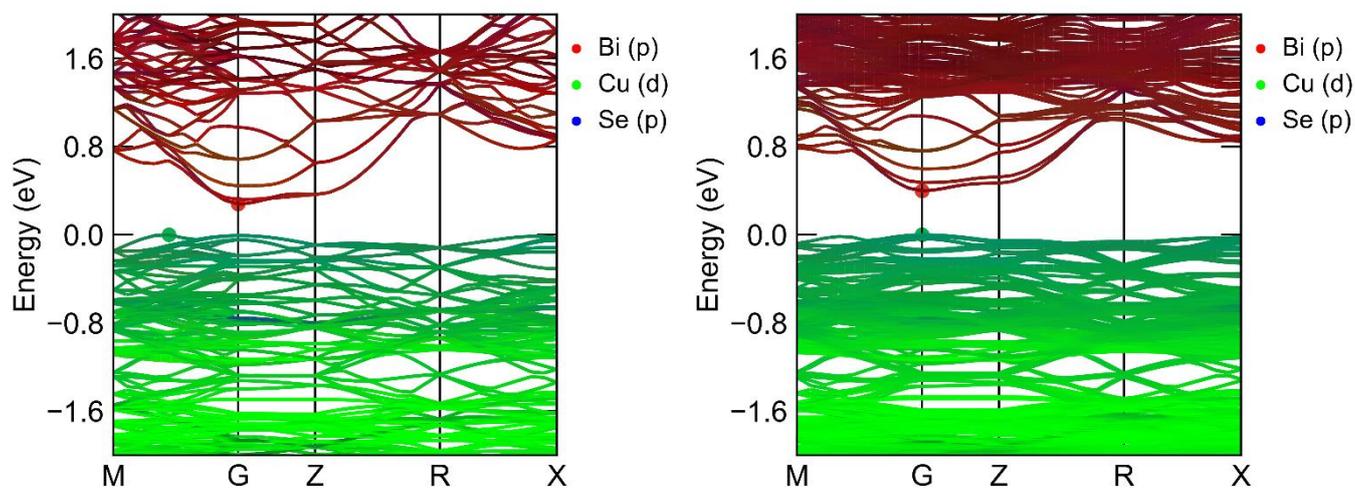

**Figure S10.** Projected band structure of (left) $Bi_{64}Cu_{64}Se_{64}O_{64}$ and (right) $Bi_{59}La_5Cu_{64}Se_{64}O_{64}$.

## Electrical and thermal transport properties

**Pristine BiCuSeO.** Transport data ($\sigma$, $\alpha$, $\kappa_{lat}$, $zT$) for pristine BiCuSeO† and BiCuSeO are presented in Fig. S11, some data from literature is also displayed.[12–14] Results for BiCuSeO† are in good consistent with previous reports: the temperature dependence of the electrical conductivity exhibits a non-degenerate behaviour; moreover, the temperature dependence of the Seebeck coefficient is also in good agreement and shows low values at $T < 600$ K as was also reported by F. Li *et al.* for BiCuSeO fabricated without ball milling during powder preparation; noteworthy values of the lattice thermal conductivity measured in parallel to SPS applied pressure direction are also close to those presented by F. Li *et al.* for both ball-milled (BM) and non-ball-milled (NBM) specimens. On the other hand, values determined in the perpendicular direction are higher by ~20%, which can be attributed to the preferential grain growth in the perpendicular direction to the axis applied during SPS uniaxial pressure as shown in Figs. S12a – d.

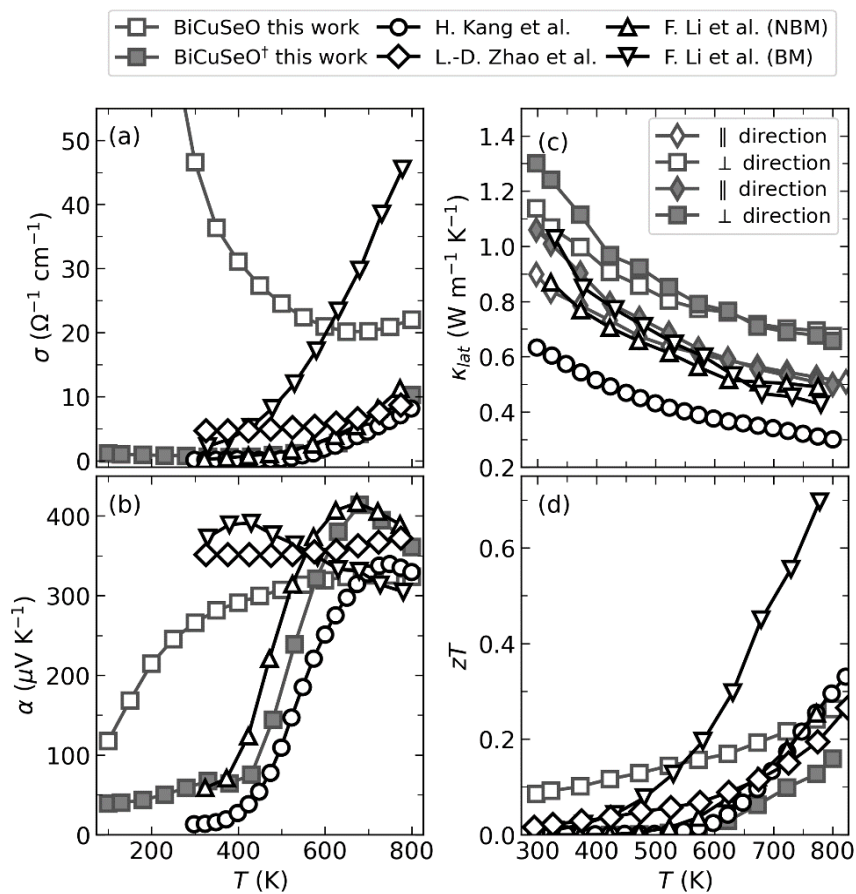

**Figure S11.** Temperature dependence of (a) the electrical conductivity, (b) the Seebeck coefficient, (c) the lattice thermal conductivity and (d) the figure of merit $zT$ for BiCuSeO†, and BiCuSeO samples. Data from previous reports are also given for comparison.[12–14]

**Anisotropy.** In order to perform all the measurements with consideration of the possible anisotropy, the samples were cut in two pieces getting cylinder-shaped and parallelepiped-shaped specimens (see Fig. S12). These two slices were used for the thermal diffusivity measurements in the parallel and the perpendicular to the SPS pressure direction, respectively (see Fig. S12e). Then the slices were cut one more time to get bars for the electrical transport properties measurements in both directions (Fig. S12e). The electrical transport properties were measured at Ioffe Institute (Saint-Petersburg, Russia) using a homemade system[15] and for some samples repeated measurements were carried out at the National University of Science and Technology MISIS (Moscow, Russia) using a laboratory-made system developed by Ltd. Cryotel (Moscow, Russia). All the measured data were in excellent agreement (see Figs. S11a and S11b).

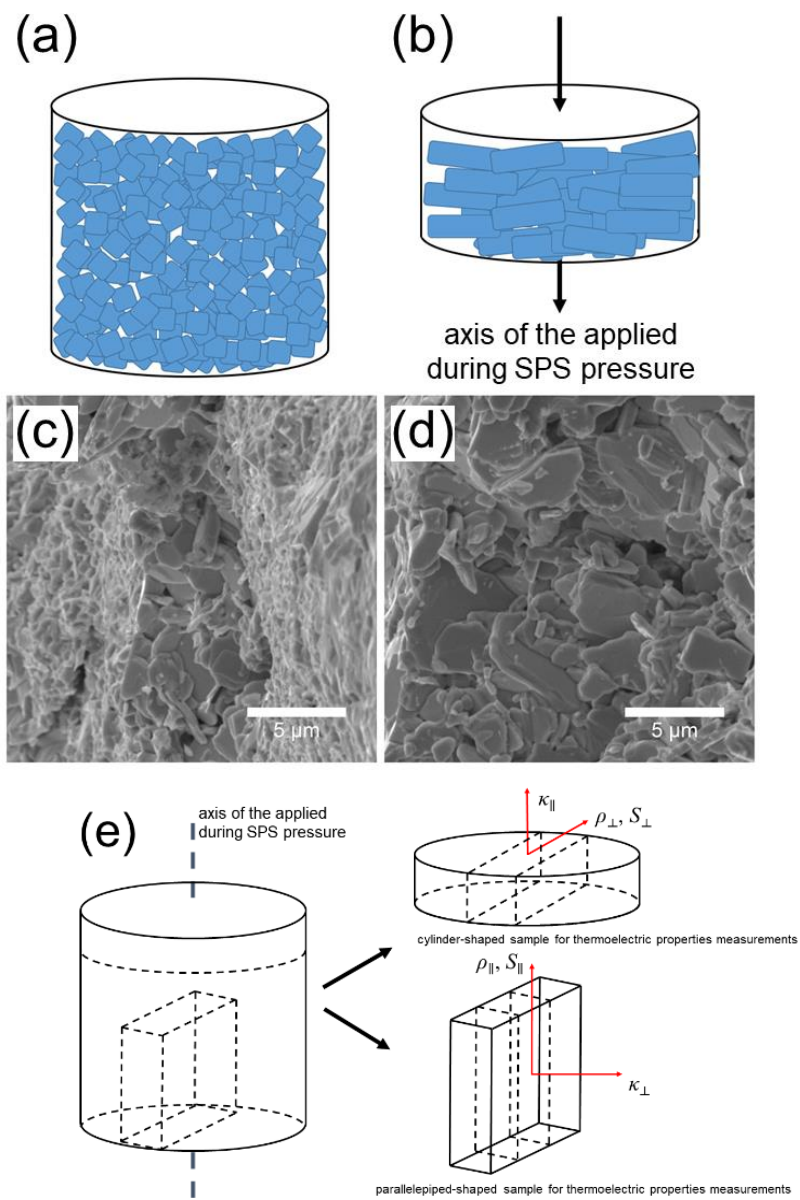

**Figure S12.** Schematic illustration of the (a) BiCuSeO powder particles before SPS and (b) the lamellar grains of the bulk BiCuSeO after SPS. SEM images of the fractured surfaces for (c) BiCuSeO and (d) $Bi_{0.96}La_{0.04}CuSeO$ bulks. (e) Schematic illustration of the measured direction for electrical and thermal transport properties.

The ratios of the electrical resistivity and the Seebeck coefficient measured in parallel and perpendicular directions do not outperform 5%, so the electrical transport properties for BiCuSeO polycrystalline specimens understudy can be assumed to be isotropic (Fig. S13). However, the thermal conductivity values indicated a difference of ~20% for those measured in the parallel and the perpendicular directions, respectively. It can be suggested that the observed anisotropy in the thermal conductivity is mainly attributed to a preferential grain growth along the perpendicular to the applied pressure direction during the SPS process (see Figs. S11c, S12a, S12b, S13e, and S13f), which is in a good consistent with the literature data and confirmed by SEM studies (see Figs. S12c, S12d).[16–18] For doped samples, the difference is decreased to 10 – 15% due to the increased electronic contribution to the total thermal conductivity.

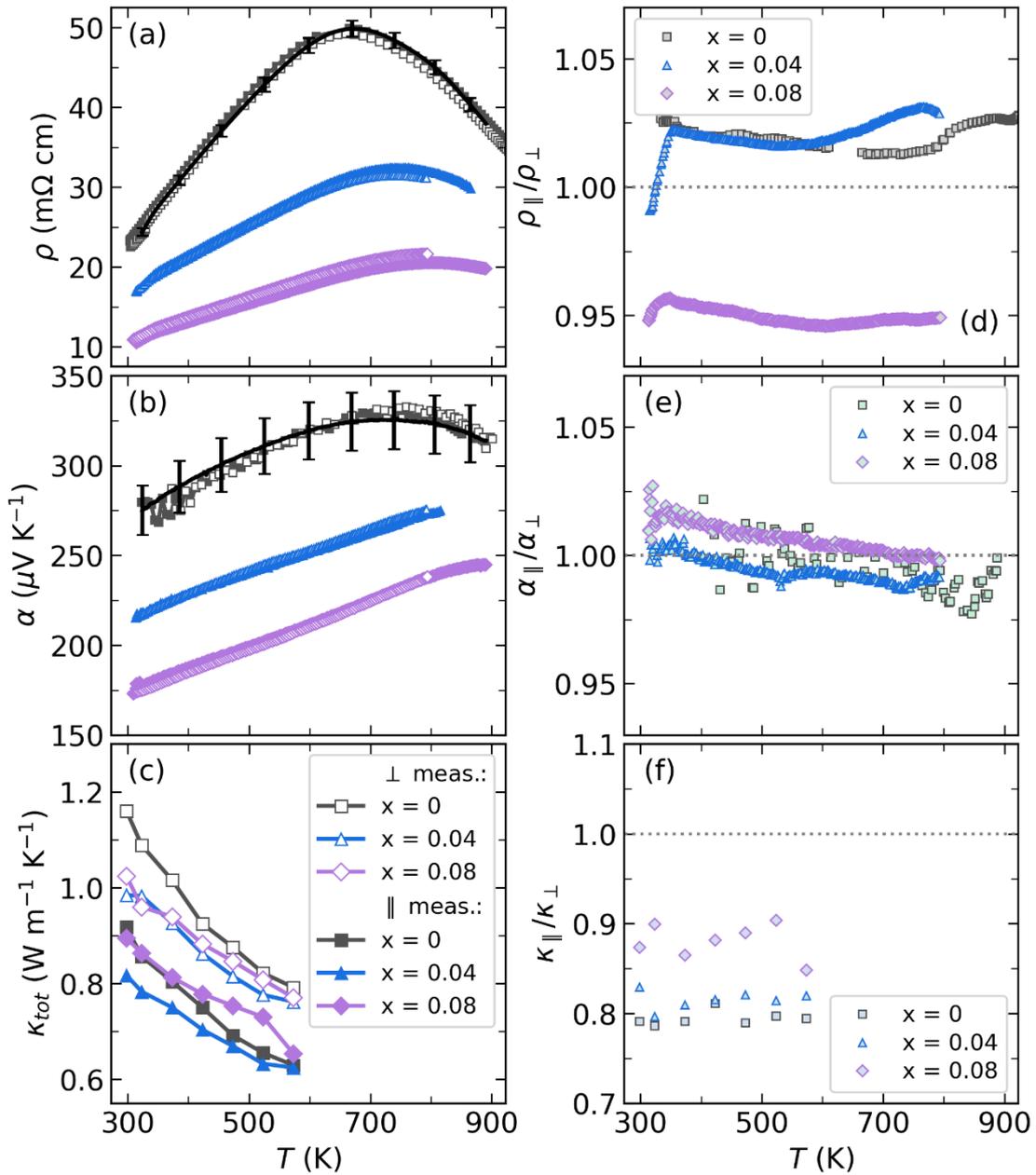

**Figure S13.** Electrical and thermal transport properties for $Bi_{1-x}Pr_xCuSeO$ ($x = 0, 0.04, 0.08$) measured in the parallel (∥) and the perpendicular (⊥) to SPS applied pressure direction. Temperature dependence of (a) the electrical resistivity, (b) the Seebeck coefficient, and (c) the total thermal conductivity. The ratios of the corresponding properties measured in different directions (d) $\rho_\parallel/\rho_\perp$ (e) $\alpha_\parallel/\alpha_\perp$, and (f) $\kappa_\parallel/\kappa_\perp$. (a, b) For $\alpha$ and $\rho$ of pristine BiCuSeO sample black solid line represents values obtained from homemade equipment at Ioffe Institute (in ∥ direction), while filled and empty squares display results obtained from homemade equipment at NUST MISIS.

## Single parabolic band model calculations

Despite the complex nonparabolic multiband electronic structure in BiCuSeO, the experimental transport data can be roughly analyzed using a single parabolic band model with relaxation time approximation. The equations shown below are valid for a single scattering mechanism where the energy dependence of the carrier relaxation time can be expressed by a simple power-law $\tau = \tau_0 E^\lambda$ ($\lambda$ is the scattering parameter).[19] The Hall carrier concentration is related to the so-called chemical carrier concentration $p$ via $p = p_H r_H$, where $r_H$ is the Hall factor, which is given by

$$r_H = \frac{3}{2} F_{1/2}(\eta) \frac{(3/2 + 2\lambda) F_{2\lambda+1/2}(\eta)}{(3/2 + \lambda)^2 F_{\lambda+1/2}^2(\eta)}, \tag{S2}$$

with the $j$-th order Fermi integrals, $F_j(\eta)$ defined by

$$F_j(\eta) = \int_0^\infty \frac{\varepsilon^j}{1 + e^{\varepsilon - \eta}} d\varepsilon, \tag{S3}$$

where $\varepsilon$ is the reduced carrier energy, $\eta$ is the reduced electrochemical potential related to Fermi energy via

$$\eta = \frac{E_v - E_F}{k_B T}. \tag{S4}$$

Here $(E_v - E_F)$ is the Fermi energy with respect to the top of the valence band, $k_B$ is the Boltzmann constant, $\lambda$ is the scattering parameter related to the energy dependence of the carrier relaxation time, $\tau$. $\eta$ values could be obtained from analysis of the Seebeck coefficient experimental data (details below).[20] For complete degeneracy ($\eta > 5$) $r_H = 1$ regardless of the scattering mechanism and $\lambda = 0$ regardless of the $\eta$ value (energy-independent charge carrier relaxation time); for nondegenerate semiconductors ($\eta < -1$), $r_H$ tends to 1.93 for ionized impurities scattering ($\lambda = 3/2$), $r_H \to 1.11$ for polar optical phonon scattering ($\lambda = 1/2$), $r_H \to 1$ for charge-neutral impurity scattering ($\lambda = 0$) and $r_H \to 1.18$ for a scattering of carries with acoustic phonons ($\lambda = -1/2$).[19–22] The values of $\eta$ and $r_H$ showed in Table S1 were calculated assuming $\lambda = -1/2$ for acoustic phonon scattering as the main charge carrier scattering mechanism. It can be clearly seen that the $R$-doping leads to a degeneration of BiCuSeO, which is in good agreement with the experimental data (see Fig. 7).

**Table S1.** Room-temperature values of the Seebeck coefficient of $Bi_{1-x}R_xCuSeO$ ($R$ = La or Pr, $x = 0 – 0.08$) samples; $\eta$, $r_H$, and $m^*$ were calculated within a single parabolic band model with acoustic phonon scattering (SPB-APS)

| Nominal composition | $\alpha$ (µV K$^{-1}$) | $\eta$ | $r_H$ |
|---|---|---|---|
| BiCuSeO† | 62 | 4.46 | 1.03 |
| BiCuSeO | 266 | –0.91 | 1.16 |
| $Bi_{0.98}La_{0.02}CuSeO$ | 243 | –0.58 | 1.15 |
| $Bi_{0.96}La_{0.04}CuSeO$ | 213 | –0.13 | 1.14 |
| $Bi_{0.94}La_{0.06}CuSeO$ | 186 | 0.32 | 1.13 |
| $Bi_{0.94}La_{0.08}CuSeO$ | 173 | 0.56 | 1.12 |
| $Bi_{0.98}Pr_{0.02}CuSeO$ | 232 | –0.42 | 1.15 |
| $Bi_{0.96}Pr_{0.04}CuSeO$ | 212 | –0.12 | 1.14 |
| $Bi_{0.94}Pr_{0.06}CuSeO$ | 173 | 0.56 | 1.12 |
| $Bi_{0.92}Pr_{0.08}CuSeO$ | 169 | 0.63 | 1.12 |

Reduced chemical potential can be calculated from:

$$\alpha = \frac{k_B}{e}\left[\frac{(\lambda+5/2)F_{\lambda+3/2}(\eta)}{(\lambda+3/2)F_{\lambda+1/2}(\eta)} - \eta\right], \tag{S5}$$

where $e$ is the electron charge.

Lorenz number can be calculated as follows:[19,20]

$$L(\eta) = \left(\frac{k_B}{e}\right)^2\left(\frac{(\lambda+7/2)F_{\lambda+5/2}(\eta)}{(\lambda+3/2)F_{\lambda+1/2}(\eta)} - \left[\frac{(\lambda+5/2)F_{\lambda+3/2}(\eta)}{(\lambda+3/2)F_{\lambda+1/2}(\eta)}\right]^2\right). \tag{S6}$$

The calculated Lorenz number of all the samples decreases with temperature, while the Hall factor exhibits similar to the Seebeck coefficient temperature dependence, as shown in Figure S14. The Hall mobility can be modeled using

$$\mu_H = \frac{e\pi\hbar^4 C_{ll}}{\sqrt{2}(k_B T)^{3/2}(m_d^*)^{5/2}\Delta_{def}^2}\frac{(3/2+2\lambda)F_{2\lambda+1/2}(\eta)}{(3/2+\lambda)F_{\lambda+1/2}(\eta)}, \tag{S7}$$

where $\hbar$ is the reduced Planck's constant, $C_{ll}$ is the elastic constant for longitudinal vibrations ($C_{ll} = dv_l^2$, where $v_l$ is the longitudinal component of sound velocity), $m_d^*$ is the density-of-states effective mass, and $\Delta_{def}$ is the deformation potential characterizing the carrier-phonon interaction (see Fig. S14c).

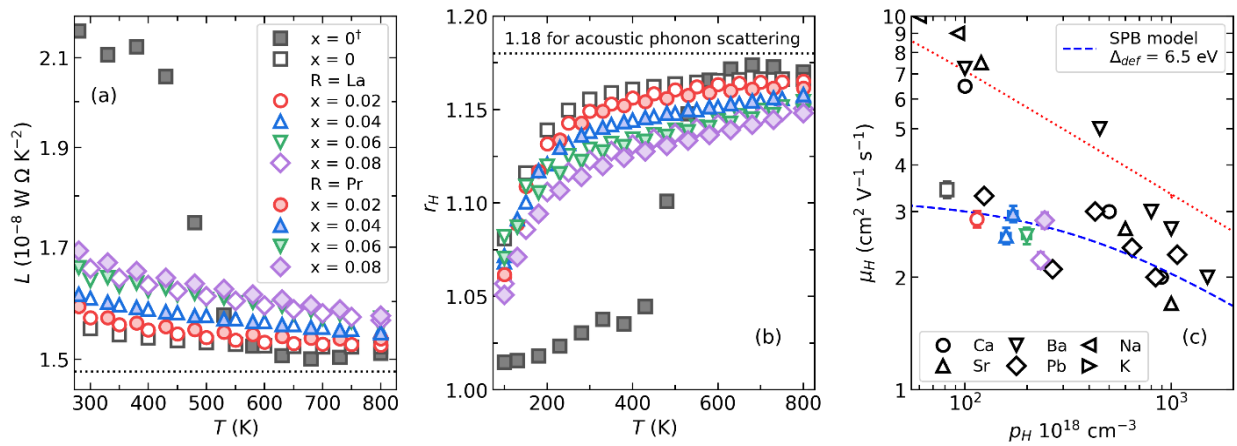

**Figure S14.** Temperature dependence of (a) the Lorenz number and (b) the Hall factor for $Bi_{1-x}R_xCuSeO$ ($R =$ La or Pr, $x = 0 - 0.08$). (c) The Hall mobility versus the Hall carrier concentration for the samples at 300 K. The experimental data (colored symbols) are compared to theoretical curves calculated under the SPB model assuming carrier mobility is limited by acoustic phonon scattering (blue dashed curve), and data from previous reports are also given for comparison.[2,5,8,23–25]

**Comparison with previous reports.** Transport properties of the $Bi_{0.96}La_{0.04}CuSeO$ and $Bi_{0.96}Pr_{0.04}CuSeO$ (not optimally doped but with close chemical composition to other reported samples) were compared with those reported for other rare-earth doped BiCuSeO.[7,12,26–30] Difference in the temperature dependence of the electrical conductivity may originate from the fabrication technique and the initial number of defects in the system. Nevertheless, in general, all the thermoelectric properties for studied samples are in good agreement with those reported for other $R$-doped BiCuSeO (even for doping with the variable-valence elements such as Yb or Sm). This is also true for the figure of merit $zT$ considering the uncertainty of its determination and possible overestimation for the cases when anisotropy of properties is not considered. Only Er- and Sm-doped samples exhibited $zT > 0.4$ at 800 K.

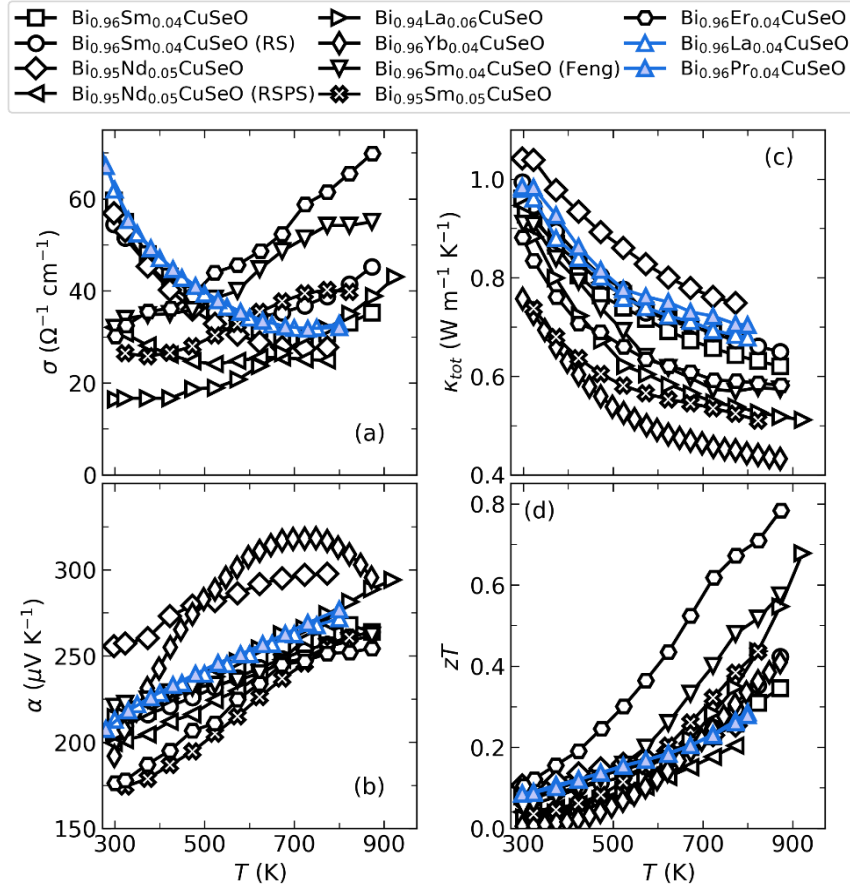

**Figure S15.** Temperature dependence of (a) the electrical conductivity, (b) the Seebeck coefficient, (c) the total thermal conductivity, and (d) the figure of merit $zT$ for $Bi_{0.96}La_{0.04}CuSeO$ and $Bi_{0.96}Pr_{0.04}CuSeO$ samples, and data from previous reports.[7,12,26–30]

**Thermal properties.** For the theoretical calculations of the lattice thermal conductivity a modified Callaway's model was used: the normal scattering process was ignored and the Umklapp scattering term is altered from $BT^3\omega^2 e^{-\theta_D/T}$ to $BT\omega^2 e^{-\theta_D/3T}$ as was suggested by Slack for materials with the temperature dependence of $\kappa_{lat} \sim T^{-1}$ at $T > \theta_D$.[31] The theoretical disorder scattering parameter was obtained using both Eqs. (9) and (10) in the main text. For the strain field fluctuation, $\varepsilon$ is directly estimated by:[32]

$$\varepsilon = \frac{2}{9}\left[(4+6.4\gamma)\frac{1+\upsilon_p}{1-\upsilon_p}\right]^2, \tag{S8}$$

where $\gamma$ is the Grüneisen parameter and $\upsilon_p$ is the Poison ratio:

$$\upsilon_p = \frac{1-2(\upsilon_t/\upsilon_l)^2}{2-2(\upsilon_t/\upsilon_l)^2} \quad \text{and} \quad \gamma = \frac{3}{2}\left(\frac{1+\upsilon_p}{2-3\upsilon_p}\right), \tag{S9}$$

with $\upsilon_t$ and $\upsilon_l$ as transverse and longitudinal sound velocities (1900 m s$^{-1}$ and 3290 m s$^{-1}$,[17] respectively).

The disorder scattering parameter of the experimental data (including both effects from dopants and vacancies) was obtained using the Klemens model.[33] This model predicts the ratio of the lattice thermal conductivities of a material containing defects ($\kappa_L^{doped}$) to that of the parent material ($\kappa_L^{pure}$) and can be written as:

$$\frac{\tan^{-1} u}{u} = \frac{\kappa_L^{doped}}{\kappa_L^{pure}} \quad \text{and} \quad u^2 = \frac{\pi^2 \theta_D \Omega \kappa_L^{pure} \Gamma_{exp}}{h v_a}, \tag{S10}$$

where $u$ is the disorder scaling parameter, $\theta_D$ is the Debye temperature, $\Omega$ is the average atomic volume, $v_a$ is the average sound velocity and $\Gamma$ is the disorder scattering parameter, respectively.

$$v_a = \left[\frac{1}{3}\left(\frac{1}{v_l^3} + \frac{2}{v_t^3}\right)\right]^{-1/3}. \tag{S11}$$

The mass-difference scattering for vacancy is given by similar to Eq. (9) in the main text expression:

$$\Phi = \frac{\sum_{i=1}^{n} c_i \left(\frac{\overline{M}_i}{\overline{\overline{M}}}\right)^2 f_i^1 f_i^{vac} \left(\frac{M_i^1 - M_{vac}}{\overline{M}_i}\right)^2}{\sum_{i=1}^{n} c_i} \quad \text{and} \quad \overline{M}_i = \sum_k \left(1 - f_i^{vac}\right) M_i^k, \tag{S12}$$

Here $f_i^1 f_i^{vac} \left(M_i^1 - M_{vac}\right)^2 = \left(1 - f_i^{vac}\right)\left(M_i^1 - \left(1 - f_i^{vac}\right) M_i^1\right)^2 + f_i^{vac} \left(0 - \left(1 - f_i^{vac}\right) M_i^1\right)^2$. However, by virial-theorem treatment for broken bonds, the above expression should be adjusted by tripling the mass difference on vacancy sites to provide more accurate calculations as proposed by R. Gurunathan *et al.*:[34]

$$f_i^1 f_i^{vac} \left(M_i^1 - M_{vac}\right)^2 = \left(1 - f_i^{vac}\right)\left(M_i^1 - \left(1 - f_i^{vac}\right) M_i^1\right)^2 + f_i^{vac} \left(-M_i^1 - 2\overline{\overline{M}}\right)^2 \tag{S13}$$

Calculated fitting coefficients are displayed in Table S2. As expected, the point defect scattering contributes to the thermal conductivity reduction by the mass and strain fluctuations. The relaxation constant $A$ increases with the doping level along with the disorder scattering parameter. It should be noted that for the BiCuSeO† sample the grain size, $L$, was assumed to be ~1000 nm, while for the samples prepared using ball milling the average grain size was assumed to be 400 nm.

**Table S2.** Calculated temperature and frequency independent fitting parameters based on Debye–Callaway model, calculated and experimental thermal conductivity for $Bi_{1-x}R_x CuSeO$ ($R$ = La or Pr, $x$ = 0 – 0.08) compounds at room temperature

| Compound | $(\Gamma + \Phi)_{calc}$ (×10⁻²) | $\Gamma_{exp}$ (×10⁻²) | $A$ (10⁻⁴¹ s³) | $B$ (10⁻¹⁵ s) | $\kappa_{lat}^{calc}$ (W m⁻¹ K⁻¹) | $\kappa_{lat}^{exp}$ (W m⁻¹ K⁻¹) |
|---|---|---|---|---|---|---|
| BiCuSeO† | – | – | 0 | 6.31 | 1.24 | 1.30 ± 0.09 |
| BiCuSeO | 2.36 | 2.58 | 0.72 | 5.37 | 1.11 | 1.14 ± 0.09 |
| $Bi_{0.98}La_{0.02}CuSeO$ | 3.19 | 6.42 | 1.30 | 5.35 | 0.97 | 0.99 ± 0.08 |
| $Bi_{0.96}La_{0.04}CuSeO$ | 4.00 | 7.88 | 1.79 | 4.83 | 0.95 | 0.95 ± 0.08 |
| $Bi_{0.94}La_{0.06}CuSeO$ | 4.82 | 8.29 | 2.53 | 4.17 | 0.93 | 0.94 ± 0.08 |
| $Bi_{0.94}La_{0.08}CuSeO$ | 5.62 | 9.08 | 4.39 | 2.96 | 0.90 | 0.92 ± 0.08 |
| $Bi_{0.98}Pr_{0.02}CuSeO$ | 3.14 | 3.91 | 1.78 | 6.57 | 1.11 | 1.08 ± 0.08 |
| $Bi_{0.96}Pr_{0.04}CuSeO$ | 3.91 | 7.54 | 2.91 | 3.64 | 0.96 | 0.96 ± 0.08 |
| $Bi_{0.94}Pr_{0.06}CuSeO$ | 4.68 | 5.84 | 0.73 | 5.95 | 1.05 | 1.01 ± 0.08 |
| $Bi_{0.92}Pr_{0.08}CuSeO$ | 5.45 | 6.80 | 3.11 | 3.52 | 0.95 | 0.98 ± 0.08 |